\documentclass[apj]{emulateapj}
\usepackage{lscape}
\usepackage{epsfig}
\usepackage{natbib}          
\shorttitle{CHANDRA OBSERVATIONS OF NGC~1600}
\shortauthors{SIVAKOFF, SARAZIN, \& CARLIN}
\slugcomment{Astrophysical Journal, accepted}

\received{2004 May 24}
\accepted{2004 Aug 19}
\begin{document}

\title{\textit{Chandra} Observations of Diffuse Gas and Luminous X-ray
Sources Around the X-ray Bright Elliptical NGC~1600}

\author{Gregory R. Sivakoff, Craig L. Sarazin, Jeffrey L. Carlin}
\affil{Department of Astronomy, University of Virginia,
P. O. Box 3818, Charlottesville, VA 22903-0818}
\email{grs8g@virginia.edu, sarazin@virginia.edu, jc4qn@virginia.edu}

\begin{abstract}
We observed the X-ray bright E3 galaxy NGC~1600 and nearby members of
the NGC~1600 group with the \textit{Chandra X-ray Observatory} ACIS S3
to study their X-ray properties. Unresolved emission dominates the
observation; however, we resolved some of the emission into 71
sources, most of which are low-mass X-ray binaries (LMXBs) associated
with NGC~1600.
Twenty-one of the sources have $L_{X} > 2 \times 10^{39}$~ergs~s$^{-1}$
(0.3--10.0 keV; assuming they are at the distance of NGC~1600)
marking them as ultraluminous X-ray point source (ULX)
candidates; we expect that only $11\pm2$ are unrelated
foreground/background sources.
NGC~1600 may have the largest number of ULX candidates in an early-type
galaxy to date; however, cosmic variance in the number of background AGN
cannot be ruled out.
The spectra and luminosity function (LF) of the
resolved sources are more consistent with sources found in other
early-type galaxies than sources found in star-forming regions of
galaxies.
The source LF and the spectrum of the unresolved emission both indicate
that there are a large number of unresolved point sources.
We propose that these sources are associated with globular clusters (GCs),
and that NGC~1600 has a large GC specific frequency.
Observations of the GC population in NGC~1600 would be very useful
to test this prediction.
Approximately 50--75\% of the unresolved flux comes from diffuse
gaseous emission.
The spectral fits, hardness ratios, and X-ray surface
brightness profile all point to two gas components.
We interpret the soft inner component ($a \la 25\arcsec$, $kT\sim
0.85$~keV) as the interstellar medium of NGC~1600 and the hotter outer
component ($a \ga 25\arcsec$, $kT \sim 1.5$~keV) as the intragroup
medium of the NGC~1600 group.
The X-ray image shows several interesting structures.
First, there is a central region of excess emission which is roughly
cospatial with H$\alpha$ and dust filaments immediately west of the center
of NGC~1600.
There appear to be holes in the X-ray emission to the north and south of the
galaxy center which are roughly coincident with the lobes of the NGC~1600
radio source.
On larger scales, there is excess emission to the northeast, which we
suggest may indicate the center of the group potential.
The group galaxy NGC~1603 shows a tail of X-ray emission to its west
which is probably due to ram-pressure stripping.
\end{abstract}


\section{Introduction}
\label{sec:intro_n1600}
X-ray emission in early-type galaxies generally comes from two sources,
hot ($kT \sim 1$ keV) interstellar gas, and hard X-ray point sources
whose properties are consistent with low-mass X-ray binaries (LMXBs).
X-ray bright galaxies (those with relatively high $L_X/L_B$ ratios, where
$L_X$ is X-ray luminosity and $L_B$ is blue optical luminosity)
are dominated by the interstellar gas
(e.g., \citealt*{FJT1985}; \citealt*{TFC1986}), while
X-ray faint galaxies (low $L_X/L_B$)
have a large proportion of emission by LMXBs
(\citealt*{FKT1994}; \citealt{P1994}; \citealt{KFM+1996}; \citealt*{SIB2000}).

Since \citet{F1989}, we have known that some of the off-nuclear X-ray point
sources in spiral and elliptical galaxies have
luminosities significantly exceeding the Eddington limit for
$1 \, M_{\odot}$.
These ultra-luminous X-ray point sources (ULXs) appear
to occur preferentially in star-forming regions,
(e.g., the Antennae: \citealt{ZF2002}); however, early-type galaxies
also contain bright point sources (e.g., NGC~1399
\citealt*{ALM2001}).
In early-type galaxies, there appear to be X-ray point sources with
$L_{X} < 2 \times 10^{39}$~erg~s$^{-1}$, consistent with accreting
objects with masses $\lesssim 15 \, M_{\odot}$ \citep*{IBA2004}.
This mass limit is in line with current estimates of the upper mass limit of
stellar mass black holes for progenitor masses $\lesssim 40 \, M_\odot$
\citep{FK2001}.
Above $2 \times 10^{39}$~erg~s$^{-1}$, the number of sources in
previously observed early-type galaxies may be consistent with the
number of expected background sources \citep{IBA2004}.
For this reason, we will adopt $L_{X} = 2 \times 10^{39}$~erg~s$^{-1}$ as
our minimum luminosity for a ULX candidate.
Although the fainter ULXs could be explained by steady,
spherically-symmetric, Eddington-limited accretion onto stellar-mass black
holes, some other mechanism is required for the brighter ULXs.

\textit{ASCA} results have indicated that the total luminosity of LMXBs in
early-type galaxies correlates better with the number of globular clusters
(GCs) than the optical luminosity of the galaxy \citep*{WSK2002}. 
\textit{Chandra} observations of early-type galaxies have also shown that a
significant fraction (20\%-70\%) of the LMXBs are associated with
globular clusters in the host galaxies
(\citealt*{SIB2000,SIB2001};
\citealt{ALM2001,KMZ+2003,SKI+2003}).
It has been suggested that most, if not all, of the LMXBs were formed in GCs
\citep{G1984,SIB2000,WSK2002},
and thus LMXBs can be used as tracers for GCs.

In this paper, we discuss \textit{Chandra} observations of NGC~1600,
an X-ray bright E3 galaxy.
NGC~1600 is the brightest member of the
NGC~1600 group; NGC~1601 ($1\farcm6$ away) and NGC~1603 ($2\farcm5$ away) are
the two nearest galaxies, both of which are non-interacting
members \citep{VVC+1992}.
We adopted the distance to NGC~1600 of 59.98~Mpc from \citet{PS1996},
which assumes $H_{0}= 75 $~km~s$^{-1}$~Mpc$^{-1}$ and uses the
\citet{FB1988} model that accounts for the Virgocentric flow and the
Great Attractor.
NGC~1600 is a boxy elliptical with a radially
anisotropic, axisymmetric three-integral distribution
function. Combined with its lack of significant rotation, its dynamics
argue for a merger origin in which the effects of gas were not very
important \citep{MG1999}.
\citet{TF2002} estimated the age of NGC~1600 to be 7.3~Gyr with
[Fe/H]$=0.41$, while \citet{TFW+2000} estimated an age of $\sim8.8$~Gyr with
[Fe/H]$\sim0.15$ through the inner $5\farcs7$ and $\sim4.6$~Gyr and
[Fe/H]$\sim0.24$ through the inner $22\farcs7$.
The colors \citep{S1973, FPM+1978} of NGC~1600 are consistent
with colors from NGC~3379, a prototypical elliptical galaxy,
suggesting star-formation has not occurred recently.
In addition to X-ray emitting gas, NGC~1600 also has cooler gas as
indicated by H$\alpha$ \citep{TS1991} and dust \citep{FPM+1999}.
Post-AGB stars seem capable of producing the necessary ionization/heating
for the H$\alpha$ and dust.

In \S~\ref{sec:obs_n1600}, we discuss the observations and data
reduction of NGC~1600. After presenting the X-ray images in
\S~\ref{sec:image_n1600}, we discuss the properties of
resolved sources in \S~\ref{sec:sources_n1600}.
The spatial distribution of the diffuse X-ray emission and structures found
in it are described in \S~\ref{sec:dif_n1600}, where they are compared to
structures in other wavebands.
We discuss the X-ray spectral properties of the sources and unresolved
emission in \S~\ref{sec:spectral_analysis_n1600}.
We estimate the gas and gravitational mass in \S~\ref{sec:mass_n1600}.
Finally, we summarize our conclusions in \S~\ref{sec:conclusion_n1600}.

\section{Observation and Data Reduction}
\label{sec:obs_n1600}

NGC~1600 was observed in two intervals (observations 4283 and 4371) on
2002 September 18--19 and September 20 with live exposures of 26,783
and 26,752~s, respectively. The ACIS-35678 chips were operated at a
temperature of $-120$~C with a frame time of 3.2~s.  We determined the
pointings so that the entire galaxy was located on the S3 chip, with
the galaxy center offset from the node boundaries of the
chip. Although a number of serendipitous sources were seen on the
other chips, the analysis in this paper is based on data from the S3
chip alone. The data were telemetered and cleaned in Very-Faint mode,
and only events with ASCA grades of 0, 2, 3, 4, and 6 were
included. Photon energies were determined using the gain file
acisD2000-08-12gainN0003.fits and corrected for time dependence of the
gain%
\footnote{See \url{http://hea-www.harvard.edu/$\sim$alexey/acis/tgain/}.}.
We excluded bad pixels, bad columns, and columns adjacent to bad
columns or chip node boundaries.

{\it Chandra} is known to encounter periods of high background
(``background flares''), which especially affect the
backside-illuminated S1 and S3 chips\footnote{See
\url{http://cxc.harvard.edu/contrib/maxim/acisbg/}\label{ftn:bkg}.}.
We determined the background count rate from the S1 chip to avoid the
enhanced flux due to the galaxy on the S3 chip.
Using Maxim Markevitch's {\sc lc\_clean}
program\footnotemark[\ref{ftn:bkg}], we found the exposure
intervals that were unaffected by background flares. The first
observation showed clear evidence of a major flare in the first 20\%
of the observation. The second observation had some small fluctuations
greater than 20\% from the mean rate. After these were filtered,
observations 4283 and 4371 had flare-free exposure times of 21,562 and
23,616~s, respectively. We created a merged events file for imaging
analysis after checking that the observations were well registered;
the separate events files were used for spectroscopic analysis. We
took the backgrounds for extended regions from the deep blank sky
backgrounds compiled by Maxim Markevitch\footnotemark[\ref{ftn:bkg}],
adjusted them to the aspect histories of our observations, and changed
the normalizations slightly to match the hard count rate
(pha=2500:3000) of the blank-sky background with a relatively
emission-free region on the S3 chip. For imaging analysis, we also
included the ``background'' due to the readout artifact in ACIS using
a script based on the {\sc make\_readout\_bg}
program\footnotemark[\ref{ftn:bkg}].

We performed the data reduction and some of the data analysis using the
{\it Chandra} analysis package {\sc ciao 2.3}%
\footnote{See \url{http://asc.harvard.edu/ciao2.3/}.}
and NASA's {\sc ftools}%
\footnote{See \url{http://heasarc.gsfc.nasa.gov/ftools/}.}.
Spectra were fit using {\sc xspec}%
\footnote{See \url{http://heasarc.gsfc.nasa.gov/docs/software/lheasoft/}.},
while correcting for the ACIS quantum efficiency (QE) degradation%
\footnote{See
\url{http://cxc.harvard.edu/cal/Acis/Cal\_prods/qeDeg/}\label{ftn:qe}.}
with the {\sc xspec acisabs} model.

\section{X-ray Image}
\label{sec:image_n1600}

\begin{figure}
\plotone{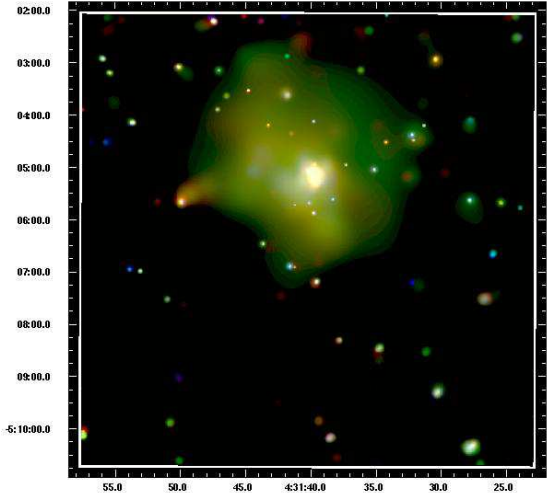}
\caption{
Adaptively smoothed {\it Chandra} true-color S3 image (with red =
0.3--1~keV, green = 1--2~keV, and blue = 2--6~keV) of NGC~1600, corrected for
exposure and background.
The total intensity scale is logarithmic and ranges from $5
\times 10^{-7}$ counts arcsec$^{-2}$ s$^{-1}$ to $1 \times 10^{-3}$
counts arcsec$^{-2}$ s$^{-1}$. The white square is the field of view
of the {\it Chandra} S3 image.
\label{fig:adaptive_n1600}}
\end{figure}

NGC~1600 has a combination of diffuse emission and resolved point
sources. In order to image the diffuse emission without deemphasizing
the point sources, we adaptively smoothed the background-subtracted,
exposure-corrected image using a minimum signal-to-noise ratio (S/N)
per smoothing beam of 3.
Figure~\ref{fig:adaptive_n1600} displays a true-color image 
of the adaptively smoothed X-ray image.
This image was created by smoothing three exposure and background-corrected
images in
soft (0.3--1~keV),
medium (1--2~keV),
and
and hard (2--6~keV) bands using the same kernel required for the
total band adaptively smoothed image, and then combining them with the
color coding red = soft, green = medium, and blue = hard.
A logarithmic intensity scale was chosen to range between total band
surface brightnesses of
$5 \times 10^{-7}$~counts~arcsec$^{-2}$~s$^{-1}$ and
$1 \times 10^{-3}$~counts~arcsec$^{-2}$ s$^{-1}$.

In Figure~\ref{fig:POSS2_n1600}, we show the corresponding Second
Palomar Sky Survey (POSS II) optical (red) image using a linear gray
scale. The field of view is the same as that in
Figure~\ref{fig:adaptive_n1600}. The largest galaxy near the upper
center is NGC~1600, and corresponds to the brightest peak in the
diffuse X-ray emission.
Another group member, NGC~1603, lies to the
east of NGC~1600.
The galaxy north of NGC~1600 is the
group member NGC~1601. On Figure~\ref{fig:POSS2_n1600}, the overlaid
regions indicate the positions of the X-ray sources discussed in
\S~\ref{sec:sources_n1600} and listed in Table~\ref{tab:src_n1600}.
All of the sources with fluxes determined to $>3 \sigma$ (shown as
squares) are clearly seen in the adaptively smoothed X-ray image
(Figure~\ref{fig:adaptive_n1600}), except for the central three
sources that are embedded in strong diffuse emission. Very few of the
weaker sources can be seen in Figure~\ref{fig:adaptive_n1600}.

The diffuse X-ray emission in Figure~\ref{fig:adaptive_n1600} shows
several interesting structures.  The central X-ray emission of
NGC~1600 is elongated in a direction which is roughly aligned with the
optical emission. However, there are some structures in the central
X-ray emission, which are discussed below
(\S~\ref{sec:dif_radial_n1600}).
The emission around the elliptical galaxy NGC~1603 in the
east appears somewhat extended, and there is a bridge of X-ray
emission extending from NGC~1603 to the west toward NGC~1600.
There is an X-ray source associated with the lenticular galaxy
NGC~1601, but it is unclear whether it is extended in this image.
On larger scales, there is a slightly elongated region of very
extended X-ray surface brightness, with excess
diffuse emission to the northeast of NGC~1600.
(Adaptively smoothed images with the sources replaced by Poisson
noise do not affect the gross morphology of the diffuse X-ray
emission; that is, these diffuse features are not due to the
smearing of point sources.)

\section{Resolved Sources}
\label{sec:sources_n1600}

\subsection{Detections}
\label{sec:src_detect_n1600}

We used the wavelet detection algorithm ({\sc ciao wavdetect} program)
with $\sqrt{2}$ scales ranging from 1 to 32 pixels to identify the discrete
X-ray source population on the ACIS S3.
Since the wavelet source detection threshold was set at $10^{-6}$,
$\la$1 false source (due to a statistical fluctuation in the
background) is expected in the entire S3 image. Source detection was
first performed on the separate observations to check their
astrometric registration; no significant offset was found. To maximize
S/N, we analyzed the wavelet detection results from the combination of
the two observations. We required our sources have wavdetect fluxes
determined at the $\ge 3 \sigma$ level for all analyses except the
identification of possible optical counterparts.

\begin{figure}
\plotone{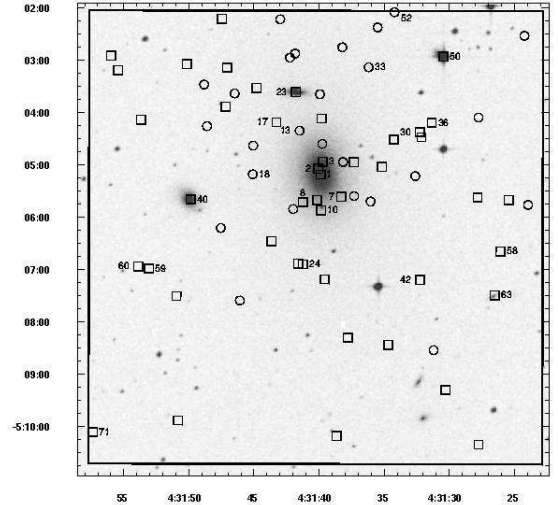}
\caption{
Linear gray scale POSS-II Red optical image of NGC~1600.
The squares and circles indicate the positions of the X-ray sources
with
S/N$>3$ and S/N$<3$ in Table~\ref{tab:src_n1600},
respectively. Sources mentioned in the text are labeled.
The dark square is the field of view of the {\it Chandra} S3 image.
\label{fig:POSS2_n1600}}
\end{figure}

At that level, the minimum detected count rate in the 0.3--6~keV band
was $2.7 \times 10^{-4}$~counts~s$^{-1}$; however, the bright diffuse
gaseous emission at the center of the galaxy makes it difficult to
establish a minimum detectable flux over the entire image.
For backgrounds of $< 0.05$~counts~pix$^{-1}$ and off-axis distances
appropriate to the S3 chip, sources with 20~counts should be
detected at a roughly uniform completeness level ($\gtrsim 85\%$)
\citep{KF2003}. For sources satisfying these criteria, the minimum
detected flux was $4.7 \times 10^{-4}$~counts~s$^{-1}$. To determine source
characteristics other than their flux, we used a local background with an
area three times that of each source's wavdetect region.
In a few cases of nearby sources, the source or background regions
initially overlapped. We slightly altered these overlapping regions,
preserving the ratio of areas and the net count rates.

We also attempted detections in multiple bands (0.3--1, 1--2,
2--6~keV) and compared detection rates to the total band (0.3--6~keV).
In these sub-bands, 45, 68, and 34\% of the total band sources were
detected, respectively. There were two, one, and five sources detected
in an individual sub-band that were not detected in the total
band. 
None of these extra detections had fluxes which were significant at the
$\ge 3 \sigma$ level.
Two, eight, and five sources were detected in only
one sub-band in addition to the total band, with two, six, and one of
those sources having total band fluxes determined at $\ge 3 \sigma$.
In this X-ray bright galaxy, performing detections by sub-bands
provided no advantage.

In Table~\ref{tab:src_n1600}, we list all discrete sources detected by
{\sc wavdetect} over the 0.3--6~keV range. The sources are ordered by
increasing projected radial distance $d$ from the center of the
galaxy. Columns 1--8 provide the source number, the IAU name, the
source position (J2000), the projected radial distance and semi-major
distance $a$ from the center NGC~1600, the wavdetect count rate with
its $1 \sigma$ error, and the S/N for the count rate. The fluxes were
corrected for exposure and the instrument PSF. The first three
sources are clearly extended;
when the count rate was determined using $1\farcs5$ circular regions
centered on the source centroid,
none of those sources are significant at the $ 3 \sigma$ level.
Although the position of Source 1 is close to the optical/IR nucleus
of NGC~1600, we are not confident it is a point source. Therefore, we
adopted the 2MASS Point Source Catalog position of R.A.\ = 4$^{\rm
h}$31$^{\rm m}$39\fs87 and Dec.\ = -5\arcdeg5\arcmin10\farcs5
as the location of the center of NGC~1600.
All listed positions include astrometry corrections based on optical/IR
counterparts (\S~\ref{sec:src_identify_n1600}); the overall absolute
astrometric errors are probably $\sim$0\farcs5 near the field center,
with larger errors further out.

In addition to the $\la$1 false source in the entire S3 field of view,
some of the detected sources may be foreground or (more likely) background
objects unrelated to NGC~1600. If we consider the fluxes of all our
detected sources, we expect $\approx$48 unrelated sources based
on the source counts in \citet{BHS+2000} and 
\citet{MCB+2000}. Using
the minimum detected $\ge 3 \sigma$ flux, we expect $\approx$15
unrelated sources.
Incompleteness will reduce both of these estimates,
especially the first number.
Unrelated sources should be spread out fairly uniformly over the S3
image (Figure~\ref{fig:adaptive_n1600}), except for the reduced
sensitivity at the center due to bright diffuse emission and at the
outer edges of the field due to reduced exposure and increased
PSF. Sources close to NGC~1600 are more likely to be associated with
the galaxy, while sources far from NGC~1600 are more likely unrelated
to NGC~1600.

\subsection{Identifications}
\label{sec:src_identify_n1600}

Sources in Table~\ref{tab:src_n1600} were cross-correlated against
optical/IR catalogs to identify possible counterparts and to improve
the absolute astrometry of the observations.
We used the Tycho-2
Catalog \citep{HFM+2000}, 2MASS%
\footnote{See
\url{http://www.ipac.caltech.edu/2mass/releases/second/doc/explsup.html}.%
\label{ftn:2MASS}}
Point Source and Extended Source Catalogs%
\footnote{When a source appeared in both 2MASS catalogs, the
Point Source Catalog positions were used.}, and the
USNO-B Catalog \citep{MLC+2003} to identify seven optical/IR
counterparts to the X-ray sources. Four of these sources were used to
determine the astrometry: Source 18 corresponds to
USNO-B1~0849-0044132, an $R=19.4$~mag object with a non-stellar PSF;
Source 23 corresponds to NGC~1601, a nearby lenticular galaxy north of
NGC~1600; Source 33 corresponds to 2MASS~04313613-0503081, a
$J=16.8$~mag star; and Source 50 corresponds to Tycho-2~4742-254-1, a
$B=11.7$~mag star. After correcting for an astrometric shift of about
$0\farcs5$, the residual astrometric errors are $\approx 0\farcs5$.

We did not use the remaining three sources for astrometry since they
were associated with extended X-ray or optical emission.
The central source (Source 1) is $\approx 0\farcs9$ from the optical
center of the galaxy (after the astrometric correction discussed above).
As the X-ray source is extended, we used a $1\farcs5$ radius circular
region centered on the X-ray source to measure or limit the flux of any
central point source;  this gave a count rate of
$(7.6 \pm 2.9) \times 10^{-4}$~counts~s$^{-1}$ ($1 \sigma$ error bars).
Since this source was neither particularly hard nor detected
very significantly as a point source, we conservatively
adopt its $3\sigma$ upper limit luminosity of
$6.7 \times 10^{39}$~erg~s$^{-1}$ as an upper
limit to a central active galactic nucleus (AGN).

Source 40 is $\approx 1\farcs7$ from the 2MASS center of NGC~1603, a
nearby elliptical galaxy east of NGC~1600. There is clearly extended
X-ray emission centered east of the source position that appears more
coincident with the galaxy center.
Since NGC~1603 extended well beyond this X-ray source
(20~mag~arcsec$^{-2}$ isophotal K fiducial elliptical aperture
semi-major axis,
r\_k20fe, of $18\farcs4$),
Source 40 may be a point source offset from the center of NGC~1603.
Similarly, Source 63 is $\approx 2\farcs9$ from
2MASX~04312667-0507309, a galaxy candidate whose r\_k20fe is
$7\farcs2$; Source 63 may also be a point source offset from this
galaxy's center.
Although Source 10 is close to an optical point source in
Figure~\ref{fig:POSS2_n1600}, the X-ray source is $\approx
3\farcs4$ south of the optical object;
thus, we do not consider this to be a reliable optical identification
for the X-ray source.

\subsection{X-ray Luminosities and Luminosity Functions}
\label{sec:src_lum_n1600}

To convert the source count rates into unabsorbed X-ray luminosities,
we used the adopted {\it Chandra} X-ray spectrum
(\S~\ref{sec:spectra_vs_res_n1600}; Table~\ref{tab:spectra_n1600},
row~3) of the resolved sources to convert 0.3--6~keV count rates into
0.3--10~keV flux.
We then assumed each source was at the distance of
NGC~1600, 59.98~Mpc,
yielding a conversion factor of $4.12 \times
10^{42}$~erg~count$^{-1}$.
Column~8 of Table~\ref{tab:src_n1600}
lists the X-ray luminosities in units of $10^{38}$~erg~s$^{-1}$, which
range roughly from $2.8 \times 10^{38}$ to $4.8 \times
10^{40}$~erg~s$^{-1}$. Since Sources 33 and 50 are likely foreground
stars, their luminosities are probably overestimates.

By restricting the sample to the sources with a uniform completeness
of 85\% ($\ge$ 20 net counts implying a count rate limit
$\ge 4.7 \times 10^{-4}$~counts~s$^{-1}$, and $d > 40 \arcsec$ corresponding
to a background $\lesssim 0.05$~counts~pix$^{-1}$,
\S~\ref{sec:src_detect_n1600}),
excluding the sources corresponding to NGC~1601 (Source 23) and
NGC~1603 (Source 40), and excluding the very bright source
corresponding to a foreground Tycho star (Source 50), we created our
analysis sample of 20 sources (see Notes in
Table~\ref{tab:src_n1600}).
We expect $11\pm2$ foreground/background sources based
on the source counts in \citet{BHS+2000} and 
\citet{MCB+2000}.

\begin{figure}
\epsfig{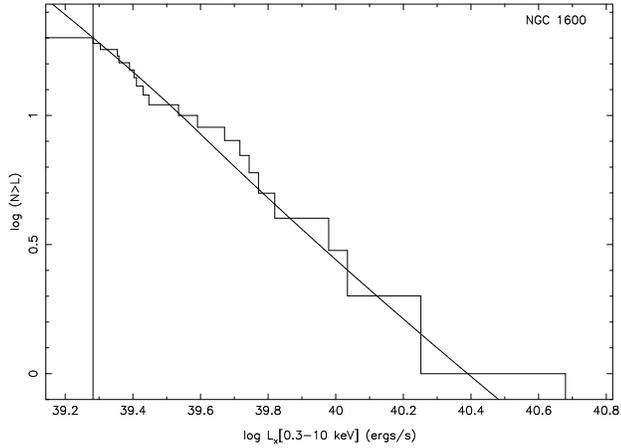}
\caption{
Histogram of the observed cumulative luminosity function of resolved
sources in our analysis sample.
The continuous curve is the sum of
the best-fit LMXB luminosity functions (eq.~\ref{eq:lfp})
and the expected background source counts.
The vertical line indicates the completeness limit of our sample.
\label{fig:lf_n1600}}
\end{figure}

In Figure~\ref{fig:lf_n1600}, we display the cumulative luminosity
function (LF) of our analysis sample. The LF should be the sum of
the point source (LMXB) population of NGC~1600 and the
foreground/background population. We fit the LF using
the same techniques we have used previously
(\citealt{SIB2000,SIB2001}; \citealt*{ISB2002}); a
single power-law, broken power-law, and a cutoff power-law were all
used to model the LMXB population. The background source population
was modeled as discussed in the previous references. A single
power-law fits the data very well 
based on the Kolmogoroff-Smirnov (KS) test:
\begin{equation} \label{eq:lfp}
  \frac{ d N }{ d L_{39} } = N_0  L_{39}^{-\alpha},
\end{equation}
where $L_{39} \equiv L_X / ( 10^{39}$~erg~s$^{-1})$. The best fit was
determined by the maximum likelihood method, and the errors (90\%
confidence interval) were determined by Monte Carlo techniques.
We
found $N_0 = 21.1^{+73.0}_{-10.0}$ and $\alpha =2.00^{+1.14}_{-0.35}$.
Although a cutoff power-law is not required by the fits, the total
luminosity of all of the sources diverges at the high
luminosity limit for the best-fit LF. The lack of very bright sources
(beyond those observed) in NGC~1600 can be explained by Poisson
fluctuations; however, it seems likely that the correct underlying LF
of the sources in NGC~1600 is either a bit steeper than the best-fit
value, or that it steepens or has a cut-off at high luminosities
beyond those observed.

Early-type galaxies tend to have broken or cutoff power-law LFs with the
break or cutoff occurring well below the luminosities measured in
NGC~1600, while star forming galaxies tend to have single power-law LFs
that extend to low luminosities
(\citealt{SIB2001,BSI2001,FJ2002,ZF2002,JCB+2003}; \citealt*{SSI2003};
\citealt*{RSI2004}). The high luminosities of the sources
in NGC~1600 make it difficult to directly compare its luminosity
function to other galaxies.
Therefore, one must either extrapolate the NGC~1600 LF down to lower
luminosities or extrapolate the LF of other galaxies to higher luminosities
in order to compare them.
The differential LF slopes at the highest
luminosities in early-type galaxies tend to be steeper than 2.5, and tend to
cluster around 1.5 in galaxies with some level of star formation.
The best-fit slope of NGC~1600 is intermediate between the two;
however, within the errors, the slope is more consistent with the
other early-type galaxies.
The normalization of NGC~1600 is higher by
at least a factor of four than in any of the galaxies in the
above references.
Extrapolating the best-fit luminosity function, we
would find $\sim$ 40 and $\sim$ 420 sources above
$5\times 10^{38}$~erg~s$^{-1}$ and $5\times 10^{37}$~erg~s$^{-1}$,
typical break luminosities and minimum observed luminosities in other
early-type galaxy observations. 
If there is a break in the luminosity function below
$2 \times 10^{39}$ erg s$^{-1}$, the numbers of fainter sources
would be reduced.

Only one of the sources in the analysis sample is fainter than $2 \times
10^{39}$~erg~s$^{-1}$.
There are six additional sources which could qualify as ULX
candidates, Sources 1--3, 7, 52, and 71. Since Sources 1--3 appear
extended, we used a $1\farcs5$ radius circular region to measure the
flux of a possible point source. The count rates of Sources 1 and 2 are
sufficient to qualify as ULX candidates; however, their fluxes are not
determined at the $3\sigma$ limit. Of the remaining candidates, only
Sources 7 and 71 are ULX candidates with well-determined fluxes.
Source 7 is not included in the analysis sample since it is only $35\farcs2$
from the nucleus of NGC~1600.
Source 71, a source at the edge of the chip did not make it into the
analysis sample due to its small number of counts, $\sim17$;
its luminosity is boosted by a large exposure correction.
Including Sources 7 and 71, we observe 21 ULX candidates.
At the flux limit corresponding to
$L_{X} = 2 \times 10^{39}$~erg~s$^{-1}$, we expect $11\pm2$
foreground/background sources based
on the source counts in \citet{BHS+2000} and 
\citet{MCB+2000} for the entire chip.
The number of ULX candidates in excess to the expected background is
$10\pm5$. Even if we consider sources with $L_{X} > 4 \times
10^{39}$~erg~s$^{-1}$, there are ten ULX candidates and only
$\approx$5 unrelated foreground/background sources are expected.
This corresponds to an excess of $5\pm3.3$%
\footnote{We use $\sqrt {N}$ statistics since they better represent the
true low-count Poisson lower limit $1\sigma$ confidence level than the
Gehrels approximation of $1+\sqrt{N+0.75}$ (Compare with eq.~[7],
eq.~[11], and Table 2 in \citet{G1986}).}
The error budgets of both
excesses are dominated by Poisson counting errors in NGC~1600.

In a sample of 28 early-type galaxies, the number of
sources with $L_{X} \ge 2 \times 10^{39}$~erg~s$^{-1}$ (0.3--10~keV )
was equivalent to the expected number of foreground/background sources
\citep{IBA2004}%
\footnote{This study used surface brightness profile distances which
are consistent with $H_{0}= 74\pm4 $~km~s$^{-1}$~Mpc$^{-1}$
\citep{TDB+2001}.}.
Although the number of ULX candidates is greater than the average
number of unrelated sources, the number of ULX candidates in NGC~1600
could be a result of cosmic variance in the foreground/background
sources. One item in support of this possibility is that the fitted
slope of the luminosity function is consistent with the typical slopes
of foreground/background luminosity functions
\citep{BAH+2001,GRT+2001}.
To examine the possibility of cosmic
variance, we compared the source densities we observed to the
0.5--2.0~keV source densities of the Chandra Deep
Fields%
\footnote{We chose to use the 0.5--2.0~keV band since all but
one of the ULX candidates were detected in soft as well as hard
bands.}.

Within two D25 ($a\le147\farcs3$), the observed source density of ULX
candidates in NGC~1600 is $2250\pm796$~deg$^{-2}$, while the remaining
field of the S3 chip has a source density of
$884\pm245$~deg$^{-2}$.
In order to compare the background source densities with those in
\citet{BAH+2001} and \citet{GRT+2001},
we converted our count rates to energy fluxes
in the 0.5--2.0~keV band using a $\Gamma=1.4$ power-law spectrum.
Then, the minimum detectable 0.5--2.0~keV flux
for our observation of NGC~1600 is
$1.36\times 10^{-15}$ erg~cm$^{-2}$~s$^{-1}$.
The Chandra Deep Fields predict
foreground/background source densities of $\sim 690\pm260$~deg$^{-2}$
(North) and $520\pm110$~deg$^{-2}$ (South).
Assuming a
background source density of $600\pm100$~deg$^{-2}$, there is an $\sim
2.1\sigma$ excess within two D25 of the galaxy corresponding to an
excess of $\sim6\pm3$ sources. The errors are dominated by counting
errors in NGC~1600 as opposed to uncertainty in the background source
density.
The excess near the galaxy favors an association of ULX
candidates with NGC~1600; however, cosmic variance cannot be ruled
out.

When one subdivides the ULX candidates into a fainter sample $L_{X} =
2$--$4 \times 10^{39}$~erg~s$^{-1}$ and a brighter sample, $L_{X} > 4
\times 10^{39}$~erg~s$^{-1}$, we find that the brighter sample source
densities are more consistent with the background than the fainter
sample. In particular, there are no sources in the bright sample,
which has better completeness than the faint sample, within two
effective radii; in the faint sample, there are four sources. One
possible explanation is that the overabundance of the bright ULX
candidates is due to cosmic variance of foreground/background sources,
while the overabundance of the faint ULX candidates is from sources
within NGC1600. The ULX candidates could then be brought more in line
with the findings of \citet{IBA2004} by a coincidence of cosmic
variance and a $\sim 40\%$ overestimate in the distance to NGC
1600. Such large differences do occasionally occur between recessional
velocity distances and surface brightness fluctuation distances;
however, in the case of NGC~1600, this would require NGC~1600 have a
peculiar velocity of $\sim+1300$km~s$^{-1}$. Since this peculiar
velocity is more typical of cluster infall and there is no nearby
large cluster, we do not think this model is likely.

If cosmic variance does not explain the excess number of bright
sources, then these bright sources are actually ULX candidates
associated with NGC~1600.
NGC~1399 has three sources with
$L_{X} \ge 2 \times 10^{39}$~erg~s$^{-1}$, the largest number at these
luminosities among previously observed early-type galaxies. NGC~1407
may have five such sources; however, its distance is highly uncertain
\citep{IAB2003}. Although \citet{JCB+2003} find that NGC~720 has
nine ULX candidates, only one has a luminosity $\ge 2 \times
10^{39}$~erg~s$^{-1}$ for $H_{0}= 75 $~km~s$^{-1}$~Mpc$^{-1}$.
Thus, we believe that NGC~1600 may have the largest number of ULX
candidates brighter than $2 \times 10^{39}$~erg~s$^{-1}$ observed in
an early-type galaxy to date.

There are a variety of models explaining the presence of ULX
sources. Some of these include favorable viewing angles of anisotropic
radiation \citep{KDW+2001}, super-Eddington accretion by high mass
X-ray binaries (HMXBs) at the thermal-timescale mass transfer
rate \citep{K2002}, accretion onto intermediate-mass black holes
\citep{CM1999}, super-Eddington accretion of LMXBs in the soft X-ray
transient state \citep{K2002}, and super-Eddington accretion from thin
accretion disks of stellar mass black holes \citep{B2002}.
If anisotropic radiation was the cause of ULXs in NGC~1600, one would
predict the existence of a large population of ULXs seen at fainter
fluxes, due to misalignment with the preferred axis of the anisotropic
radiation, in addition to the intrinsically faint and isotropic
radiating LMXB populations; our observation doesn't go deep enough to
observe such sources.

Since NGC~1600 has an observed age of 7.3 Gyr \citep{TF2002} and 
its photometric colors \citep{S1973, FPM+1978} are typical
for an elliptical galaxy, one would not expect its X-ray
binaries to be HMXBs. Comparisons of the FIR detections of NGC~1600 from
IRAS in the 60$\mu$ and 100$\mu$ bands with the radio flux (\citet{BD1985})
could be consistent with either AGN or stellar heating of interstellar dust
\citep{CB1991}.
\citet{TS1991} and \citet{FPM+1999} both found that
post-AGB stars are the likely source of the ionization of the gas and
heating of the dust in NGC~1600. Thus, the FIR detections do not
necessarily indicate that there is recent star formation as needed for HMXBs.
Additionally, the cumulative spectrum of the
ULX candidates in NGC~1600 is not consistent with the disk blackbody
model found to fit well in ULXs associated with spiral galaxies
\citep{MKM+2000}.

Intermediate-mass black holes may be created if the progenitor mass of a
star is $\gtrsim 40 \, M_\odot$ \citep{FK2001} and that star sinks to
the center of a globular cluster where it can grow up to
$\sim 10^3 M_\odot$ in $10^{10}$ years \citep{MH2002}.  Under this model,
NGC~1600 would be expected to have a large number of globular clusters.
The LMXB-GC
connection in early-type galaxies
\citep{SIB2000,SIB2001,ALM2001,KMZ+2003,SKI+2003}
means that LMXBs in the soft X-ray transient state or black hole LMXBs
with thin accretion disks would also be consistent with a large number
of globular clusters.
Thus, it seems plausible that the large population of ULXs in NGC~1600,
if they are not due to cosmic variance in the number of unrelated sources,
requires a large population of GCs.
Unfortunately, the globular cluster population of NGC~1600 does not appear
to have been determined.

\subsection{Hardness Ratios}
\label{sec:src_colors_resolved_n1600}

Hardness ratios or X-ray colors are useful for crudely characterizing
the spectral properties of sources, and can be applied to sources that
are too faint for detailed spectral analysis. We determined the
observed X-ray hardness ratios for the sources, using the same
techniques we used previously \citep{SIB2000,SIB2001,BSI2001,ISB2002}.
We define three hardness ratios as H21 $\equiv ( M - S ) / ( M + S )$,
H31 $\equiv ( H - S ) / ( H + S )$, and
H32 $\equiv ( H - M ) / ( H + M )$, where $S$, $M$, and $H$ are the
total counts in the soft (0.3--1~keV), medium (1--2~keV), and hard
(2--6~keV) bands, respectively.
As compared to our previous definitions, we have reduced the hard
band from 2--10 keV to 2--6 keV.
Since the 6--10~keV range is dominated by background
photons for most sources, this should increase the S/N of the hardness
ratio techniques. 
The hardness ratios measure observed counts, which are affected by
Galactic absorption and QE degradation in the {\it Chandra} ACIS
detectors.
In order to compare to other galaxies, it is useful to
correct the hardness ratios for these two soft X-ray absorption
effects.
Therefore, we have calculated the intrinsic
hardness ratios, denoted by a superscript 0, using a correction factor
in each band appropriate to the best-fit spectrum of the resolved
sources. The intrinsic hardness ratios and their $1 \sigma$ errors
are listed in columns 10--12 of Table~\ref{tab:src_n1600}.

\begin{figure}
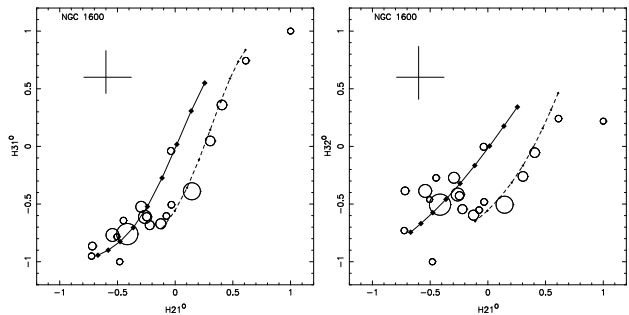

\epsfig{file=f4a.eps,angle=-90,width=0.225\textwidth,clip=}
\epsfig{file=f4b.eps,angle=-90,width=0.225\textwidth,clip=}
\caption{
Hardness ratios for the sources in our analysis sample.  Here,
H21$^{0} \equiv ( M^{0} - S^{0} ) / ( M^{0} + S^{0} )$, H31$^{0}
\equiv ( H^{0} - S^{0} ) / ( H^{0} + S^{0} )$, and H32$^{0} \equiv (
H^{0} - M^{0} ) / ( H^{0} + M^{0} )$, where $S^{0}$, $M^{0}$, and
$H^{0}$ are the counts in the soft (0.3--1~keV), medium (1--2~keV),
and hard (2--6~keV) bands, corrected for the effect of Galactic
absorption and QE degradation according to the best-fit spectra of
resolved sources. The area of each circle is proportional to the
observed number of net counts. The solid curve and large diamonds show
the hardness ratios for power-law spectral models;
the dashed curve and small diamonds show the
ratios for intrinsic absorption of $N_H = 4\times 10^{21}$ cm$^{-2}$;
the diamonds indicate values of the power-law photon number index of
$\Gamma = 0$ (upper right) to 3.2 (lower left) in increments of
0.4.
Both models underwent the same correction as the sources. The $1
\sigma$ error bars at the upper left illustrate the median of the
uncertainties.
\label{fig:colors_resolved_n1600}}
\end{figure}

Although we have plotted H31 vs.\ H21 in the past, plots of H32 vs.\
H21 allow for better separation of simple spectral models.
In Figure~\ref{fig:colors_resolved_n1600} we plot both H31$^{0}$ vs.\
H21$^{0}$ and H32$^{0}$ vs.\ H21$^{0}$ for the 20 sources in the
analysis sample. The hardness ratios for the sum of those sources are
(H21$^{0}$,H32$^{0}$,H31$^{0}$) $ = (-0.26,-0.39,-0.59)$; the
uncorrected hardness ratios are (H21,H32,H31)$ = (+0.11,-0.33,-0.24)$.
Sources with $\sim$40 net counts had errors similar to the median of
the uncertainties, $\sim0.2$. The errors scale roughly with the
inverse square root of the net counts.

In previously studied galaxies, most of the sources lie along a broad
diagonal swath extending roughly from (H21,H31) $\sim (-0.3,-0.7)$ to
$(0.4,0.5)$.
Usually, these hardness ratios were not corrected for
Galactic absorption and QE degradation; the latter effect was not
known at the time of some of the previous studies.
Since there are fewer sources and the
absorption/degradation corrections tend to push sources to the
lower-left part of the diagram, this swath is less evident in
NGC~1600.
In Figure~\ref{fig:colors_resolved_n1600}, the solid line corresponds
to hardness ratios for power-law source spectra with $\Gamma = $0--3.2.
In calculating these model hardness ratios, Galactic
absorption and QE degradation were applied to the model spectra, and
the hardness ratios were corrected for these effects using the
best-fit spectral model of the resolved sources as described above.
The dashed line corresponds to a similar model, with an intrinsic
absorbing column of $4\times 10^{21}$~cm$^{-2}$.
Most of the sources
are consistent with $\Gamma = $1.6--2.2, with the sum of these sources
corresponding to $\Gamma \approx 1.7$.
The majority of the sources lie
roughly between the two models in H31$^{0}$ vs.\ H21$^{0}$.
In H32$^{0}$ vs.\ H21$^{0}$, we see that most of the sources have an
H32$^{0} \sim -0.5$, but the two models still roughly contain the
sources. When compared to the H31$^{0}$ vs.\ H21$^{0}$ plot, the effects
of absorption in the H32$^{0}$ vs.\ H21$^{0}$ plot are larger, and the effects
of absorption and varying power-law index are more nearly orthogonal.
Although it is suggestive that the sources tend to lie on one of the two
tracks, the errors are large and the number of the sources is small, so this
could be a coincidence.

In NGC~4697 and the bulge of NGC~1291, a total of four sources had
(H21,H31) $\approx (-1,-1)$.
Scaling from the number of sources in NGC~4697, one would have only
expected $\sim$2 supersoft sources in NGC~1600, so the lack of any
strong candidates is not surprising.
Moreover, the supersoft sources in NGC~4697 would all have been below our
detection limit at the distance of NGC~1600.
Another problem is that the soft X-ray response of the {\it Chandra} ACIS
S3 detector was much worse at the time NGC~1600 was observed due to QE
degradation\footnotemark[\ref{ftn:qe}].
Although the hardness ratios are corrected for this effect, the correction
is based on the average (hard) spectrum of the sources;
this correction would be too small for supersoft sources.
Supersoft sources in NGC~1600 would have been hard to detect unless they
were very bright.

Among the sources with $> 3 \sigma$ known fluxes, we
find three sources (42, 58, and 60) with very hard spectra,
(H21$^{0}$,H31$^{0}$) $>$ $(0.5,0.5)$, and three sources (17, 36, and
50) with little hard emission, H31$^{0} \sim -1.0$.
The very hard sources
may be unrelated, strongly absorbed AGNs, similar to the sources which
produce the hard component of the X-ray background, and which appear
strongly at the faint fluxes in the deep {\it Chandra} observations of
blank fields
\citep{BHS+2000,MCB+2000,GRT+2001}; however, all three sources have large
hardness ratio errors.
The sources without hard emission may also be unrelated
foreground/background sources as indicated by studies of other
galaxies \citep{SIB2001} and deep blank sky images
\citep[e.g.,][]{GRT+2001}; Source 50 (See \S~\ref{sec:src_identify_n1600})
is clearly a foreground object, while Source 36 has large hardness
ratio errors.

The hardness ratios of Sources 1--3 are softer than a majority of the
sources in the analysis sample; however the errors of Sources 2 and 3
could bring them more in line with the analysis sample. Since these
extended sources appear softer, it is more likely that they are small
scale structures in the diffuse gas (See
\S~\ref{sec:dif_structure_n1600}), than that they are a number of
confused sources.

\subsection{Variability}
\label{sec:src_var_n1600}

With two observations we could test for variability on two time
scales, intra-observation, and inter-observation. Since the mid-points
of the observations were only $\sim 60$ ks apart, the timescales are
not remarkably different.
We searched for variability in the X-ray
emission of the resolved sources over the duration of the
{\it Chandra} observations individually and jointly using the
KS test \cite[see][]{SIB2001}. Additionally, we
compared the count rates between the two observations.
In Table~\ref{tab:src_n1600}, we report the sources that had a
$> 95$\% variability probability in any one of these four
tests.
Among the 43 $3\sigma$ flux detected sources away from the edge of the
chip, Sources 1, 8, 24, 30, and 59 were apparently variable.
Source 1 had a $2.1\sigma$ (96\% probability) rate decline in count
rate between the two observations; however, the variability was not
significantly detected within the individual observations. This
suggests that there may be a point source variable on a timescale
$>20$ ks buried in the diffuse emission of this extended source.  The
flux in a $1\farcs5$ circular region centered on the position of
Source 1, which includes the galaxy center, is nearly constant between
the two observations. This suggests that the buried variable point
source, if it exists, is not a central AGN.
For Source 8, the KS test on the joint observations yielded a 95\% probability
of variability.
Source 24 shows a variability at the 99\% probability in the first
observation; in this observation, six of its seven counts occur within
$\sim 4$ ks.
Source 30 undergoes a $2.0\sigma$ (95\% probability) rate
increase between the two observations. Finally, Source 59 has a 98\%
probability of being variable in the joint observation KS test.
With 43 sources, $\sim2$ false positives are expected at a $>
95$\% limit. Of the four sources variable at this limit, Sources 24
and 59 show the strongest behavior of variability, both in the form of
outbursts.
Although Source 71 shows 95\% variability in the first observation,
its lightcurve is irregularly sampled due to its location at the edge
of the chip. Source 13, a weakly detected source, had a very high
variability probability, 99\%, in the second observation since all
three counts in that observation occurred in the last 750 seconds.
The remaining weak sources that are marked as potentially variable did
not exhibit such clearly variable behavior as in Source 13.

\section{Unresolved X-ray Emission}
\label{sec:dif_n1600}
As mentioned in \S~\ref{sec:image_n1600}, the morphology of the
unresolved X-ray emission is complex. This emission is a combination
of unresolved point sources and diffuse gaseous emission.
Ideally, we would like to use the spectral properties of
the gas and resolved sources (See \S~\ref{sec:spectral_analysis_n1600}) to
disentangle the two as in
\citet{SIB2001} and
\citet{SSI2003};
however, 
the changing diffuse gas temperature
of NGC~1600 does not allow this
separation.  For this reason, we chose to use the entire 0.3--6~keV
band in analyzing the unresolved emission of NGC~1600.
From the spectral fits, we estimate unresolved point sources
contribute $\sim 30\%$ of the total unresolved counts.
For comparisons
with previous studies, we left in resolved sources with S/N$<3$. As
Source 1 may be the peak of the unresolved emission, we did not
exclude it.  Since Sources 2 and 3 were extended, we only excluded
regions corresponding to $1\farcs5$ circular regions located at their
centers. We used the blank-sky background to statistically remove
background events.

\subsection{Radial Profile of the Unresolved X-ray Emission}
\label{sec:dif_radial_n1600}

In order to compare the spatial distribution of the unresolved X-ray
emission in NGC~1600 with the optical emission by stars, we
adopted the Third Reference Catalogue of Bright Galaxies (RC3) values
for the optical photometry's effective radius
($r_{\rm eff}=45\farcs4$), position angle ($PA=15$), and ellipticity
($e=0.324$), which assumes a de Vaucouleurs profile \citep{VVC+1992}. The
corresponding semi-major axis, $a_{\rm eff}$ is $55\farcs2$.
Although \citet{RPD+2002} find a different profile in the infrared
(a S\'ersic profile with $n\sim 1$--2, an effective radius of
$\sim$10--15\arcsec, a semi-major axis
position angle of $\sim$10--20, and ellipticity of 0.17--0.34),
the values of the position angle and ellipticity
are roughly consistent with the optical profile.
The corresponding semi-major axis is $17\farcs6$ using the optical
ellipticity.
Since the effective radius is only used for scaling in this paper, the large
discrepancy between the two radii does not make a difference in our
study.

\begin{figure}
\plotone{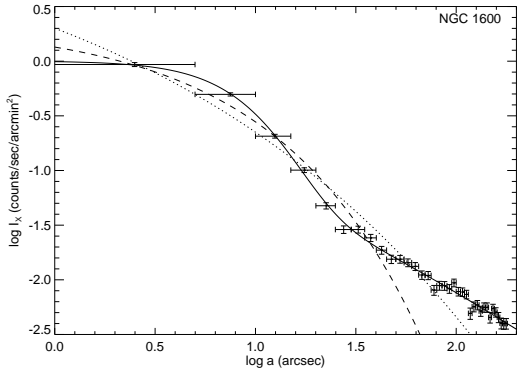}
\caption{
Surface brightness profiles, with $1 \sigma$ error bars, of the
unresolved emission (0.3--6~keV) as a function of projected semi-major
axis $a$. The dotted and dashed curves show the RC3 de Vaucouleurs
profile and J-band S\'ersic profile, respectively, with fixed effective
radii and normalizations varied to fit the X-ray surface brightness. 
The solid
curve is the best-fit double beta model profile. All fits were for $ a <
180 \arcsec$.
\label{fig:sb_n1600}}
\end{figure}

\begin{figure}
\epsfig{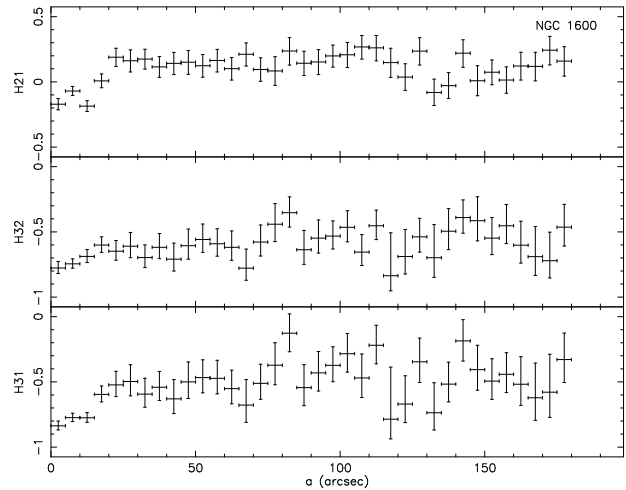}
\caption{
Hardness ratios with $1 \sigma$ error bars for the unresolved emission
as a function of semi-major axis $a$. Here,
$H21 \equiv ( M - S ) / ( M + S )$,
$H31 \equiv ( H - S ) / ( H + S )$, and
$H32 \equiv ( H - M ) / ( H + M )$, where $S$, $M$, and
$H$ are the observed counts in the soft (0.3--1~keV), medium (1--2~keV),
and hard (2--6~keV) bands.
\label{fig:colors_unresolved_n1600}}
\end{figure}

We determined the surface brightness profile
(SBP; Figure~\ref{fig:sb_n1600}) and the hardness ratios
(Figure~\ref{fig:colors_unresolved_n1600}) for a series of elliptical
annuli with semi-major widths of $5\arcsec$, extending to
$180\arcsec$.
We could have used much narrower annuli near the center of NGC~1600,
but there is significant non-axially-symmetric structure in the image there
(\S~\ref{sec:dif_structure_n1600} below).
In the SBP, the dotted line displays the (optical) best-fit de
Vaucouleurs profile with effective semi-major axis fixed at
55$\farcs$2 and the dashed line displays the (J-band) best-fit
S\'ersic $n=1.65$ profile with effective semi-major axis fixed at
17$\farcs$6.
The normalizations of the optical profiles were varied to achieve the best fit.
It is clear that neither the optical nor the infrared profile fit the
unresolved X-ray emission well.

We first tried to fit the SBP using a single beta model profile,
\begin{equation} \label{eq:beta}
  I_X(a) = I_{0}\left[1+\left(\frac{a}{a_c}\right)^2\right]^{-3\beta+1/2}
\, ,
\end{equation}
where $a_c$ is the core semi-major axis.
However, that model was clearly rejected with a $\chi^2$ of 431 for 33
degrees of freedom (dof). The best-fit single beta model
has a small core radius $a_c$, leading to essentially a power-law
model.
The model underestimates the emission in the inner 20$\arcsec$ and beyond
90$\arcsec$ while overestimating the emission between 20$\arcsec$ and
60$\arcsec$.
Figure~\ref{fig:sb_n1600} clearly shows that there are at least two
components to the X-ray SBP of NGC~1600.
Thus, we tried to fit the SBP with a double beta model:
\begin{eqnarray} \label{eq:beta+beta}
  I_X(a) &=& I_{0,{\rm{inner}}}\left[1+\left(\frac{a}{a_{c,\rm{inner}}}\right)^2\right]^{-3\beta_{\rm{inner}}+1/2} +\nonumber\\
         & & I_{0,{\rm{outer}}}\left[1+\left(\frac{a}{a_{c,\rm{outer}}}\right)^2\right]^{-3\beta_{\rm{outer}}+1/2}.
\end{eqnarray}
The core radius of the outer component $a_{c,\rm{outer}}$ is poorly
determined and consistent with zero.
That is, the outer part of the SBP could be fit by a power-law surface
brightness;
however, a pure power-law form would produce a peak in the X-ray SBP
at the very center which is not seen in the image.
Therefore, we chose to freeze the outer
core semi-major axis at $25\arcsec$.
We then found
$I_{0,{\rm{inner}}} =  0.97^{+0.08}_{-0.07}   $~counts~s$^{-1}$~arcmin$^{-2}$,
$a_{c,\rm{inner}}     =  14.4^{+3.3}_{-2.3}   $~arcsec (4.2~kpc in projection),
$\beta_{\rm{inner}}   =  1.18^{+0.33}_{-0.20} $,
$I_{0,{\rm{outer}}} =  0.039^{+0.005}_{-0.005}$~counts~s$^{-1}$~arcmin$^{-2}$,
$a_{c,\rm{outer}} \equiv 25.0                 $~arcsec (7.3~kpc in projection),
and
$\beta_{\rm{outer}}   =  0.36^{+0.01}_{-0.01} $.
The values of this fit were consistent with the error bars from
the double beta model profile with a free outer core semi-major axis, and 
the $\chi^2$ of this fit, 39.0 for 30 dof, was only 0.9 higher.
Therefore, we adopted this fit, the solid-line in
Figure~\ref{fig:colors_unresolved_n1600}, as the best-fit SBP model.
To ensure that the inclusion of Source 1 did not affect the fit, we
attempted to fit the same model without using the first annulus. The
new fit was within the error bars of the above fit.

A $\beta$ of 1.18 is much larger than in most X-ray
bright galaxies \citep{FJT1985,TFC1986}.
However, these previous fits were done with a single beta model and on
much poorer resolution data.
Below, we will argue that the inner beta model component of the SBP is
interstellar gas in NGC~1600, while the outer beta model component is
intragroup gas.
It may be that the intragroup medium has compressed the galactic gas,
decreasing the value of $a_c$ and steepening the profile (increasing
$\beta$).

Figure~\ref{fig:colors_unresolved_n1600} shows the profiles of the
hardness ratios for the unresolved emission.
As was true for the SBP, the hardness ratio profiles show a break in
behavior at $a \sim 25\arcsec$.
The inner regions of the galaxy are noticeably
softer than outer regions.
It is unlikely that this trend is due to unresolved point stellar sources.
The optical profile of the galaxy is more centrally condensed than the
X-ray SBP
(Figure~\ref{fig:colors_unresolved_n1600}), so unresolved stellar sources
should contribute more in the inner regions.
Yet, the average spectrum of the resolved sources is much harder
than any of the diffuse emission, and would not cause the central regions
of the diffuse emission to be softer.
A positive temperature gradient and/or abundance gradient
in the diffuse gas in the inner regions is the likely cause
for the softer emission;
in \S~\ref{sec:spectra_vs_unres_n1600} below, we show that the X-ray spectra
do indicate that this change in the hardness is due to a temperature
gradient.

\subsection{Structural Features in the Unresolved X-ray Emission}
\label{sec:dif_structure_n1600}

\begin{figure*}
\plotone{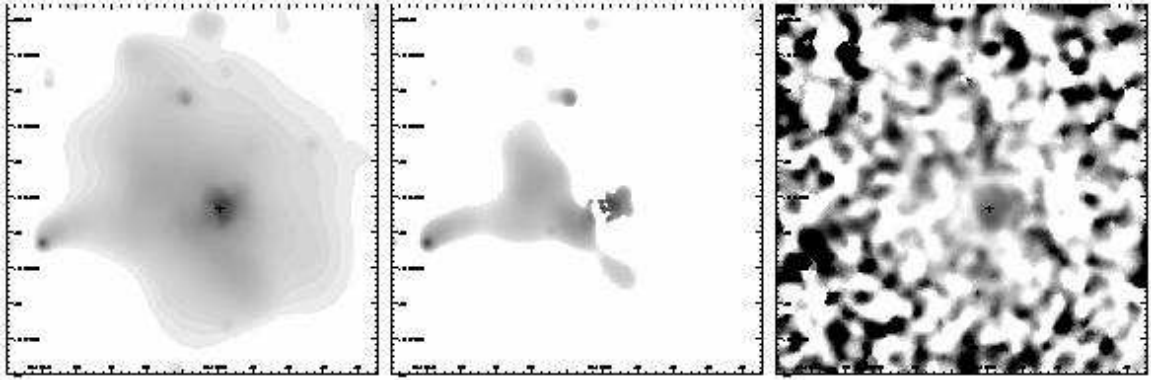}
\caption{
(Left) Adaptively smoothed {\it Chandra} S3 image (0.3--6~keV) of NGC~1600,
with sources removed and corrected for exposure and background.
The gray scale is logarithmic and ranges from
$1 \times 10^{-6}$ counts arcsec$^{-2}$ s$^{-1}$ to
$2.9 \times 10^{-4}$ counts arcsec$^{-2}$ s$^{-1}$.
(Middle) The image on the left minus the best-fit double beta model surface
brightness profile. The gray scale is also logarithmic and ranges from
$5 \times 10^{-7}$ counts arcsec$^{-2}$ s$^{-1}$
to $1.4 \times 10^{-4}$ counts arcsec$^{-2}$ s$^{-1}$.
(Right)
The image of the hardness ratio $H41 \equiv ( H + M - S ) /  ( H + M + S )$,
Gaussian smoothed with $\sigma = 5\arcsec$.
The gray scale is
linear and ranges from -0.4 (black) to 0.4 (white).
In each image, a cross marks the center of NGC~1600.
\label{fig:adapt_excess_hardness_n1600}}
\end{figure*}

In Figure~\ref{fig:adapt_excess_hardness_n1600}, we display images
where the resolved sources were replaced by the appropriate local
Poisson noise.  In each image, a cross marks the center of NGC~1600.
On the left, we display the adaptively smoothed image of the total
X-ray emission (0.3--6~keV) with logarithmic gray scale ranging from $1
\times 10^{-6}$ counts arcsec$^{-2}$ s$^{-1}$ to $2.9 \times 10^{-4}$
counts arcsec$^{-2}$ s$^{-1}$.
From this image, we subtracted the double beta elliptical SBP; the
excess emission image is shown in the middle panel with both limits of
the gray scale reduced by a factor of two. The tail extending from
NGC~1603 to the west toward NGC~1600 is shown clearly.  This image
shows that there is extended, diffuse emission around both of the
smaller galaxies NGC~1601 and NGC~1603. There is some interesting
residual structure near the center of NGC~1600, which is discussed in
more detail below (\S~\ref{sec:multilambda_n1600}). The structure near
the center includes an excess emission region to the west of the
center of NGC~1600. On larger scales, there is excess X-ray emission
to the east and northeast of NGC~1600.  In the right panel of
Figure~\ref{fig:adapt_excess_hardness_n1600}, we display the hardness
ratio image (5$\arcsec$ Gaussian smoothed) of H41 $\equiv ( H + M - S )
/ ( H + M + S)$ between -0.4 (black) and 0.4 (white). This displays
the softer central emission well. One can also see that the NGC~1603
tail appears softer than its surroundings.
These features are also evident in Figure~\ref{fig:adaptive_n1600}.

To quantify the surface brightness differences associated with the features
in Figure~\ref{fig:adapt_excess_hardness_n1600}, we determined the X-ray
surface brightness of the unresolved emission in a number of elliptical
annular pie regions.
For each annulus, we subdivided the
region by PA (Figure~\ref{fig:pa_sb_n1600}). To quantify non-uniform
morphology, we first fit a constant line to the surface brightnesses
at each annulus (the dashed-line), iteratively excluding PA regions
from the fit that diverged by more than $2\sigma$. In the innermost
annulus, emission from unresolved point sources and any inaccuracies
from using the optical isophotes will be largest.
Additionally, there
appear to be small depressions in the emission just to the north and
the south of the galaxy center, and a larger depression at the east
edge of the annulus. These make it difficult to accurately establish a
baseline for this annulus.
There is a clear excess of emission
($PA = 255\degr-345\degr, \gtrsim4.3\sigma$) to the west of the second annulus.
This excess continues in the third annulus
($PA = 255\degr-315\degr, \gtrsim3.0\sigma$), where an excess also appears
$180\deg$ away ($PA = 105-135, \sigma\gtrsim5.6$). In the fourth and
fifth annulus, the western excess has disappeared, while the eastern
excess begins to take up most of the northeast quadrant ($PA =
30\degr-120\degr, \gtrsim4.0\sigma$). 
In the fourth annulus, there is also an
isolated excess in the south ($\sim3.5\sigma$). Finally, the last
annulus clearly shows the NGC~1603 tail ($\sim4.5\sigma$).

The excess to the northeast of NGC~1600 in annuli 3--5 of
Figure~\ref{fig:pa_sb_n1600} may be due to
gas associated with the NGC~1600 group potential rather than the galaxy.
We note that this northeast excess starts at about the same radius where
the X-ray surface brightness profile in Figure~\ref{fig:sb_n1600}
has an inflection point and the larger scale component dominates.
This suggests the gas at these radii is responding to a potential with
a larger scale.
Both of these suggest that the outer gas is bound to the group, and
that the center of the group potential is to the northeast of NGC~1600.

We believe that the tail west of NGC~1603 is the result of ram pressure
stripping.
NGC~1603 is at a projected distance from NGC~1600 of $\sim$44.5~kpc
with its velocity redshifted from NGC~1600 by $\sim 284$~km~s$^{-1}$.
Since NGC~1603 is part of the NGC~1600 group, we
assumed that they were at the same distance.
Given the tail seen in the X-ray image, it is unlikely that the only
component of the velocity of NGC~1603 relative to NGC~1600 is along
the line of sight. We assume that the two transverse components of the
relative velocity are each about the same as the line of sight
component.
Thus, the total relative velocity of NGC~1603 is about
$\approx \sqrt{3} \times 284$~km~s$^{-1} \approx 490$~km~s$^{-1}$.
This would be a reasonable value for a circular orbital velocity for
NGC~1603 around NGC~1600, whose observed radial velocity dispersion
is 321~km~s$^{-1}$ \citep{FWB+1989}.
We argued above that there may be a significant potential associated with
the NGC~1600 group, which would increase the estimated velocity.

We estimated the gas densities and pressures in NGC~1603 and in its
environment to see if ram pressure could be sufficient to strip gas
and form the tail.  For the gas in NGC~1603, we assumed three uniform
density spherical annuli with widths of $2\arcsec$ each. Since the
X-ray emission from NGC~1603 is softer than the center of NGC~1600, we
assume a spectrum with $kT=0.6$~keV and solar abundances for the gas
in this galaxy.
With these assumptions, we find that the electron
number densities of the gas in NGC~1603 are $4.9\times10^{-2}$,
$1.9\times10^{-2}$, and $5.4\times10^{-3}$ cm$^{-3}$ and the pressures,
$P_{\rm gas}$,
are $9.4\times10^{-11}$, $3.6\times10^{-11}$, and $1.0\times10^{-11}$
dyne cm$^{-2}$ in the 0--2, 2--4, and 4--6\arcsec\  annuli.  We
estimated the density in the group gas around NGC~1603 from the X-ray
surface brightness in a hemispherical annulus centered on NGC~1603
(radius from NGC~1600 = 6$\arcsec$--12$\arcsec$,
$PA = 0^\circ-180^\circ$).  This surface brightness agrees within the
errors with the prediction by the modeled surface brightness profile
at the semi-major distance of NGC~1603 from NGC~1600.  We assumed a
spectrum with $kT=1.5$~keV and solar abundances. This gave an ambient
gas electron density of $4.0\times10^{-3}$ cm$^{-3}$ at the projected
semi-major distance of NGC~1603.
Assuming a relative velocity of
490~km~s$^{-1}$, the ram pressure of the ambient gas would be
$P_{\rm ram} \approx 1.6\times10^{-11}$ dyne cm$^{-2}$. The ram
pressure would be lower if NGC~1603 is at a larger radius than its
projected radius (i.e., in front of or behind the center of the
NGC~1600 group). Roughly speaking, the condition for ram pressure to
strip the gas is that $P_{\rm ram} > P_{\rm gas}$, which appears plausible in
the outer 4--6$\arcsec$ of NGC 1603.
Since the tail points to the west towards the
higher density gas near NGC~1600, the ambient gas density may have
been higher in the past, making ram-pressure stripping easier
than it is at its present position.

The NGC~1603 tail has a softer spectrum 
(Figure~\ref{fig:adapt_excess_hardness_n1600} right)
than the surrounding gas.
Hardness ratios from counts in the same pie annuli used to construct
Figure~\ref{fig:pa_sb_n1600} confirm that the tail is softer in H21
by $\approx 2.6\sigma$.
Thus, the jump in emission due to the tail is dominated by a jump in
soft emission, consistent with ram-pressure stripping of cooler galactic
gas by hotter group gas.

NGC~1601 also has a velocity redshifted from NGC~1600 by $\sim
309$~km~s$^{-1}$, and its projected distance from NGC~1600 is only
$\sim$28.5~kpc. Since the emission associated with NGC~1601 is
fainter, its estimated thermal pressure is smaller than in NGC~1603.
This, combined with its slightly higher velocity, suggests that
ram-pressure stripping should be stronger in NGC~1601 compared to
NGC~1603.
Although there is a hint of a tail toward the east of
NGC~1601 in Figure~\ref{fig:adapt_excess_hardness_n1600},
it has only a $1.3\sigma$ significance when the surface brightness in
a hemispherical annulus east of NGC~1601 is compared to a similar
annulus west of NGC~1601.
There are a number of possible explanations
of why we do not see a significant tail around NGC~1601 while we see
one for NGC~1603.
First, NGC~1601 might have a small transverse velocity; the tail would
then be projected onto NGC~1601 along our line of sight.
Additionally, the
tail could be too faint, the projected velocity could be smaller than
we estimated for NGC~1603, or projection effects could mean NGC~1601
is actually in a lower density region in the group gas than we
calculate from the surface brightness profile and its projected radius.

\subsection{Central X-ray Structure and Multi-Wavelength Comparisons}
\label{sec:multilambda_n1600}

\begin{figure}
\plotone{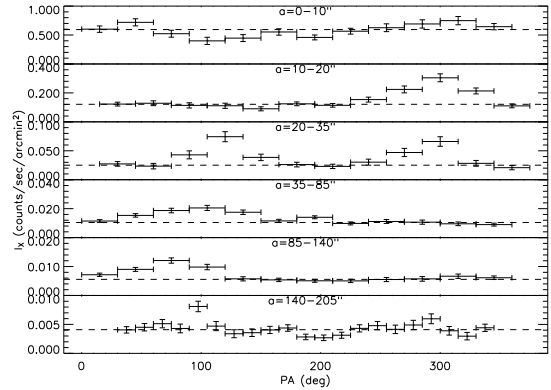}
\caption{
Surface brightness profiles, with $1 \sigma$ error bars, of the
unresolved emission (0.3--6~keV) as a function of PA for annuli with
varying projected semi-major radii $a$.  The dashed line indicates the
best-fit constant surface brightnesses of each annulus, iteratively
excluding PA regions more than $2\sigma$ from it.  The last annuli
excludes PA regions that extend beyond the S3 chip.
\label{fig:pa_sb_n1600}}
\end{figure}

The X-ray emission near the center of the galaxy is elongated
similarly to the optical emission (Figures~\ref{fig:adaptive_n1600}
and \ref{fig:POSS2_n1600}). However, the details of the central X-ray
structure are complex.
Figures~\ref{fig:adapt_excess_hardness_n1600} and
\ref{fig:pa_sb_n1600} display the excess emission immediately west of the
galaxy center, as well as holes in the emission immediately north and
south of the galaxy center.

\begin{figure}
\plotone{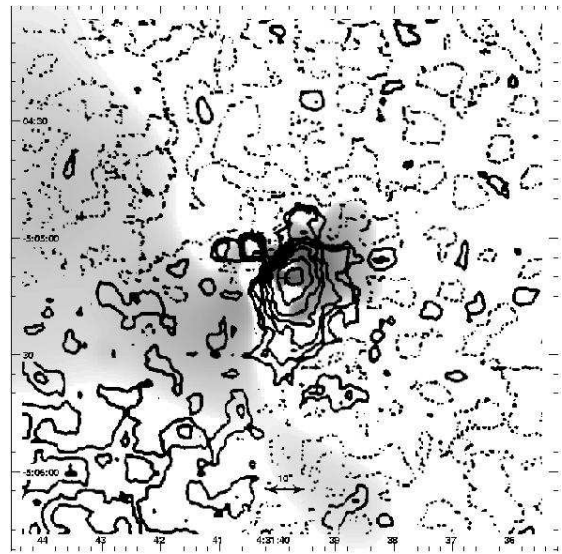}
\caption{
H$\alpha$ +[\ion{N}{2}] contours \citep{TS1991} overlaid on
adaptively smoothed image of unresolved X-ray emission ($140
\times 140 \arcsec$) with a best-fit elliptical model for the surface
brightness profile removed. The contours indicate line emission of 0,
1.7, 2, 4.1, 6.2, 8.3, 12.4, 18.6, 27.6, 41.4 $\times 10^{-17}$ erg
cm$^{-2}$ arcsec$^{-2}$.  The cross indicates the center of the
galaxy.
\label{fig:halpha_n1600}}
\end{figure}

One of the purposes of this {\it Chandra} observation was to compare
the X-ray structure to the extended emission-line filaments and dust.
In Figure~\ref{fig:halpha_n1600}, we
display the H$\alpha$ +[\ion{N}{2}] image \citep{TS1991} overlaid on
the excess emission image of the inner $140 \times 140 \arcsec$.
The peak of the H$\alpha$ corresponds to excess emission west of the
galaxy center; however, the detailed structure of the H$\alpha$ and
excess X-ray emission do not have a one-to-one correspondence.
\citet{MPC+1996} find $\sim 5\times5\arcsec$ of
H$\alpha$+[\ion{N}{2}] emission centered on the optical center, with a
slight north-south elongation.
There is excess X-ray emission centered on the galaxy; however, a
correlation of excess emission is made difficult by the possible
presence of a central AGN and holes in the X-ray emission $\sim 2\farcs5$
to the north and south of the galaxy center.

\begin{figure}
\plotone{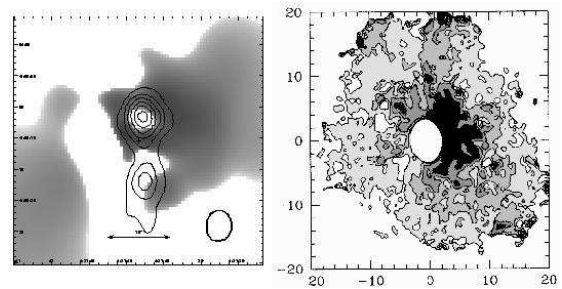}
\caption{
Left, adaptively smoothed gray scale image of unresolved X-ray
emission ($40 \times 40 \arcsec$) with the best-fit
beta+power-law surface brightness profile removed. Contours are radio
(4.885 GHz: 1, 2, 3, 4, 5, 6 mJy) with the beam in the lower-right corner
\citep{BD1985}.
The cross indicates the center of the galaxy.
Right, gray scale image of $V-R$ (0.625, 0.63, 0.635, 0.64);
the darker (redder in color) values in correspond to larger $A_V$
\citep{FPM+1999}.  
\label{fig:dust+radio_n1600}}
\end{figure}

On the left of Figure~\ref{fig:dust+radio_n1600}, we display the
excess emission image of the inner $40 \times 40 \arcsec$.
It is overlaid by 4.885 GHz radio contours from \citet{BD1985}.
The double radio lobe structure is oriented in the north-south direction, and
the lobe positions are roughly consistent with holes in the
X-ray emission. Although the southern lobe's centroid is a bit offset from
the hole, the $5\arcsec$ Gaussian smoothed image of H41 indicates that
the lobe is cospatial with a region of softer emission. Considering
the smoothing scale, this could be consistent with cool gas
surrounding a radio bubble. No variation in H41 is observed near the
northern lobe. 
These holes in X-ray emission may be due to the radio lobes,
although this is uncertain.

In Figure~\ref{fig:dust+radio_n1600}, we also display a $V-R$
reddening map \citep{FPM+1999}. The darker areas of the color index
map correspond to regions of larger $V-R$ and extinction $A_V$. There
is a clear filamentary structure to the west corresponding to a mean
$A_V = 0.034\pm0.030$ \citep{FPM+1999}%
, where the standard deviation is due to non-uniform extinction over
the area of measurement.
The optical extinction can be converted into a hydrogen column density assuming
$N_H = 5.9 \times 10^{21} A_V/R_v$~cm$^{-2}$\citep{S1978}.
Taking $R_V \equiv A_V / ( B - V )  = 3.2$, there is an excess $N_H$ of
$0.63 \pm 0.55 \times 10^{20}$~cm$^{-2}$ beyond the Galactic
value of $4.86 \times 10^{20}$~cm$^{-2}$.
If one assumes that the X-ray emission of a mekal model with $T=0.85$~keV
and solar abundance is absorbed by this extra column, the surface brightness
would be reduced by only $\sim$2\% compared to that expected with no excess
absorption.
Of course, we actually observe an excess in the X-ray emission in this region of
slightly larger extinction.
The excess emission to the west of the galaxy is $\sim $1.5--2.5 times
the values expected from the surface brightness profile model.

Both the dust lanes and emission line filament appear to be cospatial
in projection with the enhanced X-ray emission.
There are several possible explanations for such a correlation.
It might be that the hot and cool gas are in thermal contact, and that
heat is being conducted from the hot gas to the cool gas \citep{S1992}.
This might cool the X-ray gas;
if it remained at nearly the same pressure due to the weight and pressure
of the surrounding hot gas, the density and X-ray emissivity of the X-ray
gas would increase.
Mixing between the hot and cool gas could have a similar effect, as long
as the X-ray gas didn't mix to a cool temperature out of the X-ray band.
Third, the cool gas might result from radiative cooling of X-ray gas.
The cooling timescale at the center of NGC~1600 is on the order of 300
Myr,  which is consistent with this explanation.
On the other hand, it might be difficult to understand the presence of
dust grains in cooled X-ray gas.

\section{Spectral Analysis}
\label{sec:spectral_analysis_n1600}

We extracted spectra of the resolved sources and diffuse emission in
NGC~1600, restricting analysis to the 0.7--9~keV range. The lower
limit was taken to avoid calibration uncertainties, while there are
few non-background counts beyond 9~keV.
Since the telescope collecting area changes very rapidly near 2~keV and
calibration problems led to poor fits in that spectral
region, we chose to excise the 1.9--2.1~keV band when we fit line
models (mekal).
We do not find that this edge significantly affected
fits of continuum models. All of the spectra were grouped to have at
least 25 counts per spectral bin prior to background correction to
enable our use of $\chi^2$ statistics. 
In some cases, the fitted range of the spectra did not extend up to
9 keV because there were too few counts to form a bin up to this
limiting energy.

The results of the spectral fits are summarized in
Table~\ref{tab:spectra_n1600}. Spectra were extracted for the
resolved point sources (`Sources') and the unresolved diffuse emission
excluding the point sources (`Unresolved'). The third column gives the
geometric region for the spectrum; `Field' implies the entire S3 chip.
The value of the absorbing column density ($N_H$) applied to all
components of the model emission spectrum is given in column 4.  In
this and other columns, values in parentheses are fixed (not allowed
to vary). The fixed value of $N_H$ is the Galactic value from
\citet{DL1990}. To correct for the QE degradation, we used the
{\sc xspec acisabs} model. Other elliptical galaxies are known to have
both hard (point source) and soft (diffuse gas) components
\citep{SIB2001}. Under the hard component, we fit three different
models, thermal bremsstrahlung (`bremss'), power-law (`power'), and
disk blackbody (`diskbb').
We always used the mekal model for the
emission spectrum from hot diffuse gas.
For the hard component,
columns 5--7 give the spectral model(s), the temperature $T_h$ (for
bremsstrahlung or disk blackbody) or photon number spectral index
$\Gamma$, and the unabsorbed flux of the hard component(s), $F^h_X$
(0.3--10~keV). Similarly, columns 8--10 give the temperature $T_s$,
overall heavy element abundance relative to solar, and flux for the
soft mekal component(s). For the unresolved emission, the spectra
exclude regions around each of the $> 3\sigma$ flux determined
resolved sources. The last two columns give the number of net counts
in each spectrum, and $\chi^2$ per dof for the best-fit model. All
errors reported in the spectral analysis are 90\% confidence level
errors.
Brackets are used when either the upper or lower bound on the confidence
interval was unconstrained.

The background spectrum for the resolved sources were determined
locally, using the same nearby regions as discussed in
\S~\ref{sec:sources_n1600}. For the spectra of the unresolved
emission, we used the deep blank sky backgrounds compiled by Maxim
Markevitch\footnotemark[\ref{ftn:bkg}].

The spectra of several spatial regions were analyzed. We have adopted
the ellipticity and PA of the optical de Vaucouleurs fit. Some of the
regions are scaled to the elliptical optical isophote containing
one-half of the optical light, the ``one effective radius'' region.
In NGC~1600, the semi-major axis of this isophote, $a_{\rm eff}$, is
$55\farcs2$. Since there were few source counts near the center of
the galaxy, we fit the sources for the entire field. We searched for
changes in unresolved emission with radius using annuli containing
$\sim$ 1000 net counts.

\subsection{X-ray Spectrum of Resolved Sources}
\label{sec:spectra_vs_res_n1600}

\begin{figure}
\epsfig{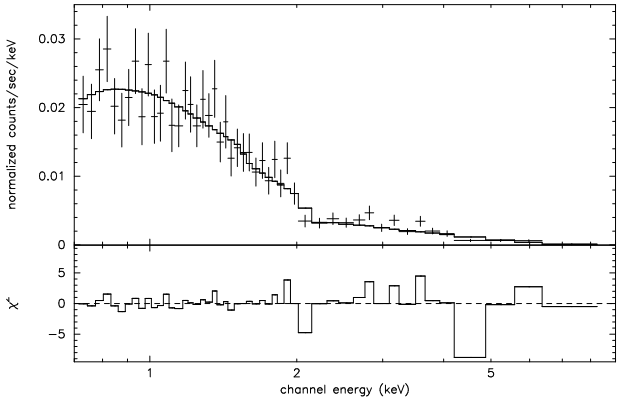}
\caption{
Upper panel: the cumulative X-ray spectrum of the resolved
sources with $>3 \sigma$ detected flux in the entire field of
NGC~1600.  The spectrum has $1 \sigma$ error bars and is overlaid by the
solid histogram of the best-fit model spectrum
(Table~\ref{tab:spectra_n1600}). Lower panel: the contribution
to $\chi^2$ with the sign indicating the sign of the residual.
\label{fig:src_spec_n1600}}
\end{figure}

All resolved source spectral analysis was performed on sources with
$> 3\sigma$ determined fluxes. Sources 1--3 were excluded because we are
unsure that they are point sources.
Sources 23 (NGC 1601), 40
(NGC 1603), and Source 50 (GSC 04742-00254) were also excluded because they
are clearly not associated with NGC~1600. Since there are few source
counts near the galaxy's center, we first extracted the cumulative
spectrum of resolved sources within three effective radii, yielding
only 372 net counts. This was insufficient to produce a
well-constrained fit, so we chose to fit the sources in the entire
field.
The observed spectrum, containing 1318 net counts, is shown in
Figure~\ref{fig:src_spec_n1600}.

The combined spectrum of the sources
was reasonably well-fit by either a thermal bremsstrahlung model with
$kT_h = 4.73$~keV (Table~\ref{tab:spectra_n1600}, row 1) or power-law
model with a photon number spectral index of $\Gamma = 1.76$ (row 3).
The fits were not improved significantly when the absorbing column was
allowed to vary (rows 2 and 4), so we fixed the hydrogen column at the
Galactic value, $N_H = 4.86 \times 10^{20}$ cm$^{-2}$ \citep{DL1990}.
Since the sources were very luminous, we also attempted a disk
blackbody model; however, its $\chi^{2}$ was worse by $\sim
20$ for the same number of dof. Finally, we attempted a disk blackbody
+ power-law spectral model ($kT_{disk} = 1.44$, $\Gamma = 1.95$,
power-law responsible for $\sim70\%$ of flux). This model had a
$\chi^{2}$ lower than our bremsstrahlung or power-law fits by $\sim2$;
however, the dof was also reduced by 2. Additionally, this fit had an
unconstrained disk temperature, and poorly constrained power-law. We
could have adopted either the bremsstrahlung or the power-law model
for our best-fit. For comparison to other papers and since the
power-law was constrained more tightly than the bremsstrahlung temperature,
we adopted the power-law model as
our best-fit. This model, including Galactic absorption and QE
degradation, is shown in Figure~\ref{fig:src_spec_n1600} with the  
residuals to the fit. The power-law index is consistent
with what was found from the hardness ratios. It is softer than the
best-fit value of the sources simultaneously fit in a survey of 15
early-type galaxies
\citep{IAB2003}; however, it is consistent with both the spread of
indices found when the galaxies are fit separately and the softer
indices of the higher luminosity subsamples.

Approximately half of the sources in the entire field are
expected to be foreground/background sources. On the other hand only
$\sim 1$ is expected to be a foreground/background source within three
effective radii.
To ensure that unrelated sources do not heavily bias the
spectral fits, we compared fits to the less constrained three
effective radii spectrum with the fits to the entire field spectrum.
Since all of the fits were consistent within the errors of the entire
field spectrum fits, we do not believe the foreground/background sources
heavily bias the spectral fits.

\phn

\phn

\subsection{X-ray Spectra of Unresolved Emission}
\label{sec:spectra_vs_unres_n1600}

\subsubsection{Projected Spectra}
\label{sec:spectra_vs_unres_proj_n1600}

\begin{figure}
\epsfig{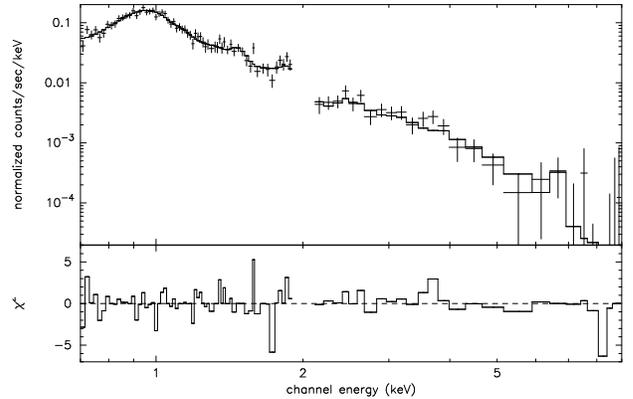}
\caption{
Upper panel, the cumulative X-ray spectrum of the unresolved emission
within 1 $a_{\rm eff}$ of NGC~1600.
The notation is the same as in
Figure~\protect\ref{fig:src_spec_n1600}.
\label{fig:unres_spec_n1600}}
\end{figure}

For the unresolved emission, the spectrum of the inner effective
radius had 3662 net counts. This spectrum is shown in
Figure~\ref{fig:unres_spec_n1600}. First, we attempted to model the
unresolved emission with a soft mekal component representing the
emission by diffuse interstellar gas (Table~\ref{tab:spectra_n1600},
row 11). The $\chi^2$ was large, 168.6 for 87 dof. Since the
unresolved emission includes unresolved point sources as well as
diffuse gas, we added the adopted best-fit spectrum of the resolved
sources to model the unresolved sources. The fit was much improved,
$\chi^2 = 110.0$ for 86 dof; however, it was still rejected at the
$>95\%$ level. We allowed the absorbing column to vary, but this did
not significantly improve the fit and was still consistent with the
Galactic value.
Therefore, we have assumed Galactic absorption
for the remaining fits. Since the hardness ratios of the unresolved
emission indicated spectral evolution at $a \sim 25\arcsec$, we
attempted a two-temperature gas solution (row 9). We found a good fit,
$\chi^2 = 78.7$ for 83 dof; however, the abundance of the low
temperature gas was unconstrained. Therefore, we tried a two-temperature gas
model with the abundances tied together.
This fit, with a low mekal
temperature of $0.85\pm 0.04$~keV, a high mekal temperature of
$2.55^{+0.52}_{-0.86}$~keV, and an abundance of $1.07^{+1.00}_{-0.40}$
solar, was almost as good as when both abundances were free. We also
attempted a mekal cooling flow model; however, its fit,
$\chi^2 = 94.4$ for 85 dof, was worse than the two-temperature
model. Therefore, we adopted the two-temperature gas model (row 10) as our
best-fit model for unresolved emission
(Figure~\ref{fig:unres_spec_n1600}).
Although the best-fit flux of this model
from the unresolved point sources was not large, the upper
limit on the flux of $1.41 \times 10^{-13}$ erg cm$^{-2}$ s$^{-1}$
in 0.3--10~keV band still indicates unresolved sources could be a non-trivial
source of emission in the inner effective radius.

Since this galaxy was bright enough to fit multiple annuli of
unresolved emission, and there were indications of multiple temperatures of
the diffuse gas, we attempted to determine the radial dependence of
gas temperature and abundance. We used eight annuli each with approximately
1000 net counts, out to $a=180\arcsec$. In
Table~\ref{tab:spectra_n1600} rows 11--18, we show the results of
assuming the model was the sum of the resolved point source model with
its normalization free and a single temperature gas mekal model. Most of
the fits produced reasonable $\chi^2$. Any radial changes in the
abundance are dwarfed by the errors of the fits; although the best-fit
abundances were mainly subsolar.
On the other hand, it
is clear that the first two annuli have a much lower temperature,
$\sim 0.85$~keV, than the outer five annuli, $\sim 1.5$~keV. The
temperature of the third annulus is intermediate between the two
temperatures.  These fits are in rough agreement with the two-temperature
model found within one effective radius, $a < 55\farcs2$;
however, the flux of the unresolved sources is smaller and the
temperature fit for the hotter gas is larger when fitting the
inner effective radius at once, as opposed to in multiple annuli.
The best-fit normalization of the power-law and mekal model suggest that
for $a<180\arcsec$ they both contribute approximately equally to
the flux. Within $1 a_{\rm eff}$, the diffuse gas is dominant by at least
two-to-one.
From the best-fit fluxes of the annular fits, we can estimate the
X-ray luminosity (0.3--10~keV) to be $\sim 2.9 \times 10^{41}$~erg~s$^{-1}$
in gas and $\sim 2.4 \times 10^{41}$~erg~s$^{-1}$ in unresolved sources for
$a<180\arcsec$. Around 30--40\% of the gaseous luminosity comes from the
cooler gas. The resolved $3\sigma$ sources in the entire field have an X-ray
luminosity of $\sim 1.4 \times 10^{41}$~erg~s$^{-1}$.
In $a<180\arcsec$, this
scales to $\sim 0.4 \times 10^{41}$~erg~s$^{-1}$.
Bolometric corrections
increase the source luminosity by 44\% and the
total gaseous luminosity by 28\%.

Using the RC3 optical profile,
$a<180\arcsec$ corresponds to $\sim80\%$ of the total optical light,
$L_{B,80\%} = 8.5\times 10^{10}$ $L_{B\odot}$ \citep*{OFP2001}.
The source X-ray-to-optical ratio is
$3.3 \times 10^{30}$~erg~s$^{-1}$~$L_{B\odot}^{-1}$,
approximately four times that found in NGC~4697 \citep{SIB2001} or NGC~1553
\citep{BSI2001}. It is more than ten times the expected contribution from
discrete sources reported in \citet{OFP2001}.
Since the unresolved sources dominate the total source luminosity, there is
an excess in the flux of faint sources in addition to the excess in the number
of detected sources. It is unlikely that cosmic variance would lead to an
excess number of sources at all fluxes.
This suggests that the number of LMXBs
found in NGC~1600 is not proportional to the number of stars as
estimated by the optical light, or that the unresolved source flux is
overestimated.
The most likely solution is that NGC1600 has a high specific frequency
of globular clusters; this is discussed in more detail later in the paper
(See \S~\ref{sec:conclusion_n1600}).

The large unresolved hard component flux in the spectral models
suggests that there are many unresolved point sources.
This is also roughly consistent with the observed luminosity
function for resolved sources
(\S~\ref{sec:src_lum_n1600}).
If we extend the best-fit luminosity function down to
$10^{36}$~erg~s$^{-1}$ and adjust the normalizations to account for
the observed sources within $3 a_{\rm eff}$, we expect an unresolved
flux of $\sim 2.2 \times 10^{-13}$ erg cm$^{-2}$ s$^{-1}$ in the
0.3--10~keV band. Although this is approximately half of the flux
estimated by the spectral fits, it is within the errors of the
spectral fit flux.
This agreement between the luminosity function and
spectral fits could point to a large number of unresolved point
sources.
It might also suggest that there is no strong break in the luminosity
function at lower X-ray luminosities than observed.

\subsubsection{Deprojected Spectra}
\label{sec:spectra_vs_unres_deproj_n1600}

We also fit deprojected spectra.
We did the deprojections by fitting the spectra from the outside annulus
to the inside.
Each annulus was fit assuming the best fit model for the emission in
each of the outer annuli, and assuming simple geometric projection to
include the emission from outer annuli in the inner ones.
First, we allowed the power-law normalization and mekal model to vary
within each annulus.
The outer four annuli had unphysically large abundances, and
in general the abundances were poorly constrained.
The temperature results were similar to those for the projected fits.
Next, we fit the deprojected spectra assuming a solar abundance in all
annuli to ensure that the abundance problems in the previous
deprojection did not cause problems in the temperature determination.
This deprojection also produced results similar to the projected fits,
except the jump between a low temperature gas and a high temperature
gas moved inward.
Since these results are qualitatively the same as the projected fit,
but we had to assume a constant abundance, we choose not to present
the details of these results.

\subsubsection{Galaxy and Group Gas}
\label{sec:spectra_vs_unres_galgroup_n1600}

Both the unresolved emission within one effective radius and the
unresolved emission in annuli out to $180\arcsec$ point to a model
involving two components of hot gas.
In the inner 25--40$\arcsec$
region, there is gas with $kT \sim 0.85$~keV.
With a stellar velocity dispersion of 321~km~s$^{-1}$
\citep{FWB+1989}, NGC 1600 has a stellar kinetic temperature of
$\sim0.65$~keV, predicting an X-ray temperature of $\sim1\pm0.2$~keV
based on \citet{DW1996,BB1998}. This prediction is consistent with
the measured X-ray temperature.
At semi-major distances beyond 40$\arcsec$, the gas is hotter,
$kT \sim 1.5$ keV.
This temperature is more consistent with the gas temperature of
X-ray bright groups \citep{HP2000} than X-ray luminous early-type
galaxies \citep{OPC2003}. 
Since the abundances are poorly constrained,
we cannot tell if there is any gradient in the abundance, or if the
average abundance is higher or lower than solar.

Near the same radius where there is a transition in spectrum, we saw
transitions in the behavior of the surface brightness profile and the
hardness ratios.
This break in behavior of the surface brightness, hardness ratio, and
spectrum at $a \sim$ 25--40$\arcsec$ (7.3--11.7 kpc in projection)
might be explained if emission from a lower temperature diffuse gas from the
interstellar medium (ISM) of NGC~1600 dominates in the inner regions and
a higher temperature diffuse gas from the intergalactic medium (IGM)
of the NGC~1600 group provides the emission at larger radii.

The total gas X-ray-to-optical ratio is
$3.4 \times 10^{30}$~erg~s$^{-1}$~$L_{B\odot}^{-1}$;
the galaxy gas luminosity is $\sim$30--40\% of the total gas luminosity.
Compared to the $L_{X, \rm{bol}}$--$L_{B}$ relationships from
\citet{OFP2001}, the galaxy and the total gaseous luminosity are
consistent with the standard deviation around the best-fit
relationship to the early-type galaxies excluding AGNs, brightest
cluster galaxies and dwarfs. The total emission is consistent with the
best-fit relationship for the brightest group galaxies.

\section{Mass Determination}
\label{sec:mass_n1600}

By combining the SBP model and the spectral fits of the diffuse gas,
we estimated the gas and gravitational mass around NGC~1600.
In eq.~(\ref{eq:beta+beta}), the SBP is represented as the sum
of two functions.
Because the X-ray emissivity is a quadratic function of the density,
eq.~(\ref{eq:beta+beta}) does not lead to a simple expression for
the gas density as the sum of two terms.
Thus, we assumed that the two terms in the SBP represented two cospatial
phases of gas, a 0.85~keV gas following the
inner beta model profile and a 1.5~keV gas following the outer beta model
profile.
To remove the $\sim 30\%$ contribution to the counts by unresolved
point sources (See \S~\ref{sec:dif_n1600}), we multiplied the SBP by $0.7$.
We integrated the sum of the physical densities over volume to determine the
gas mass, and
we used hydrostatic equilibrium to estimate the gravitational mass.
For both estimates, we assumed the galaxy was shaped as
an oblate spheroid.

\begin{figure}[b]
\plotone{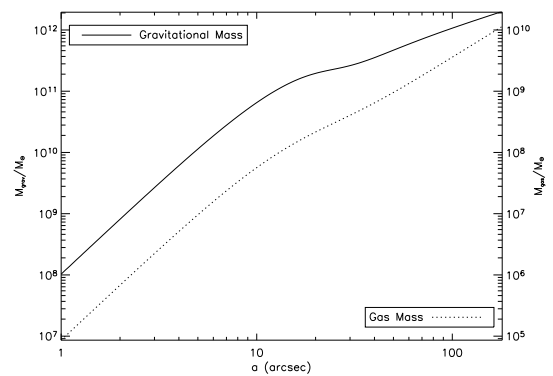}
\caption{
The estimated gas and gravitational mass profiles around NGC~1600.
The dotted line and right axis display the gas mass profile, while
the solid line and left axis display the gravitational mass profile.
\label{fig:mass_n1600}}
\end{figure}

In Figure~\ref{fig:mass_n1600}, we display the resulting gravitational and
gas mass profiles.
At $a\sim 20 \arcsec$, there is a slope change in the
gravitational mass profile. This indicates the likely presence of two
gravitational potentials, supporting the view that the exterior gas is group
gas and the interior gas is galaxy gas.
On the other hand,
the total gravitational mass within $a<180 \arcsec$ is only
$2.0\pm0.5 \times 10^{12} M_\odot$ with errors in the hot gas temperature
dominating the error budget.
The mass-to-light ratio for $a<180 \arcsec$
is $M / L_B = (24\pm6) \, M_\odot/L_{B\odot}$.
Within one effective radius, the mass-to-light ratio is 
$ ( 10 \pm 3 ) \, M_\odot/L_{B\odot}$.
The total mass is not that much larger than might be expected for NGC~1600
alone, which may not be consistent with the suggestion that the outer
component is due to a group dark matter potential.

The total gas mass within $a<180 \arcsec$ is $1.14\pm0.06 \times 10^{10}
M_\odot$, with errors in the surface brightness dominating the error
budget.
Assuming a normal elliptical galaxy stellar mass loss rate per stellar
luminosity of
$1.3\ \times 10^{-11} M_\odot/L_{B\odot}$ \citep{MB2003}, the rate of
stellar mass loss for the entire galaxy is about $1.4 \, M_\odot$~yr$^{-1}$.
Almost all of the gas mass around NGC~1600 can be attributed to
stellar mass loss in NGC~1600, assuming an age of 7.3~Gyr
\citep{TF2002}.
This is not surprising since NGC~1600 dominates the stellar population
of the NGC~1600 group.
On the other hand, it is not consistent with a large amount of
intergalactic group gas in the system.

\section{Conclusions}
\label{sec:conclusion_n1600}

We have used \textit{Chandra} observations to study the X-ray emission
from the point sources and unresolved emission of NGC~1600.
Since NGC~1600 is $\sim 60$~Mpc away, we could only resolve the very
brightest point sources. We detected 71 total point sources, of which
45 were bright enough to have fluxes determined at $\ge 3\sigma$.
We identified two of the sources with foreground stars,
two with known galaxy members of the NGC~1600 group,
and three with non-stellar objects, presumably galaxies.
We did not clearly detect a point source at the center of NGC~1600,
and we put a conservative 0.3--10~keV upper flux limit of
$6.7 \times 10^{39}$~erg~s$^{-1}$ on the luminosity of the central AGN.

Twenty-one of the sources without clear
associations to objects besides NGC~1600 are brighter than $2 \times
10^{39}$~erg~s$^{-1}$ (0.3--10~keV); approximately $11\pm2$ of those
sources are expected to be unrelated foreground/background
sources.
The excess is strongest within two D25, where we observe a source
density of $2250\pm796$~deg$^{-2}$ compared to the expected
$600\pm100$~deg$^{-2}$. 
NGC~1600 may have the largest number of ULX
candidates observed to date in an early-type galaxy,
although cosmic variance in the background source population cannot
be completely ruled out.
A combination of cosmic variance for the high flux sources and a
distance overestimate of $\sim40\%$ could also reduce the number of
ULX candidates; however, this model requires NGC~1600 have a peculiar
velocity of $\sim+1300$km~s$^{-1}$.
As found in
\citet{IAB2003}, the bright LMXBs have softer spectra than is typical
for fainter LMXBs in early-types galaxies.  The spectrum of sources in
NGC~1600, $\Gamma=1.76^{+0.10}_{-0.09}$, and the luminosity function
slope, $\alpha =2.00^{+1.14}_{-0.35}$, both suggest that these sources
are different than ULX candidates observed in star-forming galaxies.
The age of the galaxy and the X-ray spectra of the sources argue against
these sources being HMXBs.

Since this
galaxy is X-ray bright, we expected the unresolved emission to
dominate the flux. Although this is true, the contribution from
unresolved point sources is nearly as large as that due to the diffuse gas in
the best-fit models.
Even if one uses the lower limit on the
unresolved point source flux, unresolved point sources are responsible
for $\ga$25\% of the unresolved flux.
Combined with the large normalization of the X-ray luminosity function,
this suggests that NGC~1600 has a larger population of LMXBs than found in
most previously observed galaxies
\citep{SIB2001,BSI2001,FJ2002,ZF2002,JCB+2003, SSI2003, RSI2004}.
When normalized to the optical luminosity of the galaxy, the
source X-ray-to-optical ratio is also very high,
$L_X / L_B = 3.3 \times 10^{30}$~erg~s$^{-1}$~$L_{B\odot}^{-1}$.
This ratio is a factor of two larger than found in most early-type galaxies.
Recently, \citet{G2004} suggested that total X-ray source luminosities
correlate better with near-infrared luminosities than with optical
luminosities. From the extrapolated 2MASS $K_s$ magnitude (8.04),
$M_{K_{s},\odot} = 3.39$, and the distance of NGC~1600, we find that
the source X-ray-to-near-infrared ratio is almost an order of
magnitude higher than the average found in \citet{G2004}.
\citet{G2004} limited the sources he considered in his total X-ray
luminosities to $L_X > 10^{37}$ erg~s$^{-1}$, so part of the difference
might be explained if NGC~1600 has a very large number of fainter sources.

Previous studies have shown that a large fraction of LMXBs in elliptical
galaxies are located in globular clusters
\citep{SIB2000,SIB2001,ALM2001,KMZ+2003,SKI+2003},
and that the luminosity of LMXBs may correlate better with the GC
population than the optical luminosity
\citep{WSK2002,SKI+2003}. 
Unfortunately, we were unable to find a determination of the GC population
in NGC~1600, which is more distant than most galaxies which have GC
observations.
However, one explanation for the large number of LMXBs in NGC~1600 would
be that this galaxy has a large GC specific frequency.
Thus, we predict that NGC~1600 will be found to have large numbers of GCs.
Observations of the GC population in NGC~1600 would be very useful
to test this prediction.
Such a large population of GCs might be connected to NGC~1600's position as
the central elliptical galaxy in a group with a significant potential;
it may be more closely related to cD galaxies, which have larger GC
populations
\citep[e.g.,][]{H1991} than normal ellipticals.

We note that many of the bright X-ray sources associated with NGC~1600 are
at large distances from the galaxy center.
One possibility is that these sources are due to cosmic variance in the
background source population. For the sources within two D25, the variance
would have to be much larger than the $\sim 8\%$ cosmic variance
between the Chandra Deep Fields found by \citep{RTG+2002}.
Another possibility is that these sources are actually located in intergalactic
GCs in the NGC~1600 group, rather than being directly associated with
NGC~1600.
Recently, intergalactic GCs have probably been detected in nearby groups
and clusters
\citep[e.g.,][]{BCF+2003}.

The X-ray image, surface brightness profile, and spatially resolved
spectra suggest that there are two components to the gas around NGC~1600.
There is a component which is centered on NGC~1600, which has a small
spatial scale ($\la 25\arcsec$) and a lower temperature (0.85 keV),
which we propose is gas which is bound to NGC~1600.
A second component is centered to the northeast of NGC~1600, has a larger
spatial scale, and is hotter (1.5 keV);
we argue that this gas is bound to the dark matter potential of the
NGC~1600 group.
The X-ray image suggests that
the center of the potential of the group is slightly displaced
from the center of NGC~1600.
One would expect that NGC~1600 would be moving in the group potential, and
this motion could affect the distribution of the galaxy and group gas.

We also observe structure to the diffuse emission on small
scales. Excess emission is seen west of the center of NGC~1600 in the inner
$35\arcsec$. This excess emission is partially cospatial with H$\alpha$
and dust filaments.
Possible models to explain this correlation of hot and cooler gas include
thermal conduction between the two phases, radiative cooling of the hot X-ray
gas to form the cooler gas, and mixing of the hot and cool gas.
Directly to the north and south of the galaxy center, there
are holes in the X-ray emission.
These holes are roughly coincident with the lobes of the radio source,
suggesting that there may be two radio bubbles being blown in the hot gas.
Finally, we see that NGC~1603, a galaxy east of NGC~1600, has a tail of
soft X-ray emission to its west.
Calculations indicate that this is likely to be a ram-pressure stripped tail
of ISM from NGC~1603, removed as a result of motions of this galaxy
through the surrounding group gas.
\acknowledgements

Support for this work was provided by the National Aeronautics and Space
Administration through $Chandra$ Award
Numbers
GO2-3100X,
GO2-3099X,
GO3-4099X,
AR3-4005X,
GO4-5093X,
and
AR4-5008X,
issued by the $Chandra$ X-ray Observatory Center, which is operated by the
Smithsonian Astrophysical Observatory for and on behalf of NASA under
contract NAS8-39073. 
GRS acknowledges the receipt of an Achievement Reward for College Scientists
fellowship.
Partial support was also provided by the Celerity Foundation.
This research has made use of the NASA/IPAC Extragalactic Database (NED) which
is operated by the Jet Propulsion Laboratory, California Institute of
Technology, under contract with the National Aeronautics and Space
Administration, the SIMBAD database, operated at CDS, Strasbourg, France, and
the Digitized Sky Surveys, produced at the Space Telescope Science Institute
under U.S. Government grant NAG W-2166.

\bibliography{ms}

\clearpage

\LongTables
\begin{landscape}
\tabletypesize{\scriptsize}
\pagestyle{empty}
\begin{deluxetable}{lccccrrrrcccl}
\tabletypesize{\scriptsize}
\tablecaption{Discrete X-ray Sources in NGC~1600}
\tablehead{
Src.& & R.A.& Dec.&
\multicolumn{1}{c}{$d$}&
\multicolumn{1}{c}{$a$}&
\multicolumn{1}{c}{Count Rate}&
&
&
&
&
&
\\
No.&
Name&
(h m s)&
($\arcdeg$ $\arcmin$ $\arcsec$)&
\multicolumn{1}{c}{($\arcsec$)}&
\multicolumn{1}{c}{($\arcsec$)}&
\multicolumn{1}{c}{($10^{-4}$ s$^{-1}$)}&
\multicolumn{1}{c}{S/N}&
\multicolumn{1}{c}{$L_X$}&
H21$^{0}$&
H31$^{0}$&
H32$^{0}$&
Notes \\
(1)&
(2)&
(3)&
\multicolumn{1}{c}{(4)}&
\multicolumn{1}{c}{(5)}&
\multicolumn{1}{c}{(6)}&
\multicolumn{1}{c}{(7)}&
\multicolumn{1}{c}{(8)}&
\multicolumn{1}{c}{(9)}&
(10)&
(11)&
(12)&
(13)}
\label{tab:src_n1600}
\startdata
 1&CXOU J043139.8$-$050511&04 31 39.88&$-$05 05 11.4&\phn\phn0.9&\phn\phn1.0&   \phn38.91$\pm$3.10&   12.55&         160.3&$-0.50^{+0.11}_{-0.09}$&$-0.96^{+0.09}_{-0.03}$&$-0.88^{+0.22}_{-0.08}$&a,b,d,e\\
 2&CXOU J043140.0$-$050504&04 31 40.05&$-$05 05 04.7&\phn\phn6.4&\phn\phn6.5&   \phn15.00$\pm$1.94&\phn7.75&      \phn61.8&$-0.29^{+0.25}_{-0.21}$&$-0.88^{+0.24}_{-0.09}$&$-0.78^{+0.36}_{-0.15}$&a,b\\
 3&CXOU J043139.7$-$050456&04 31 39.71&$-$05 04 56.6&   \phn14.1&   \phn15.5&\phn\phn7.31$\pm$1.42&\phn5.16&      \phn30.1&$-0.48^{+0.28}_{-0.21}$&$-0.91^{+0.43}_{-0.08}$&$-0.77^{+0.77}_{-0.20}$&a,b\\
 4&CXOU J043138.1$-$050456&04 31 38.15&$-$05 04 56.9&   \phn29.0&   \phn42.3&\phn\phn1.78$\pm$0.70&\phn2.55&   \phn\phn7.3&$-0.80^{+1.09}_{-0.19}$&$-0.67^{+0.63}_{-0.25}$&$+0.27^{+0.67}_{-1.11}$&\nodata\\
 5&CXOU J043140.1$-$050540&04 31 40.15&$-$05 05 40.7&   \phn30.6&   \phn33.2&\phn\phn4.26$\pm$1.05&\phn4.05&      \phn17.5&$-0.78^{+0.66}_{-0.18}$&$-0.48^{+0.36}_{-0.25}$&$+0.47^{+0.42}_{-0.87}$&\nodata\\
 6&CXOU J043139.7$-$050436&04 31 39.75&$-$05 04 36.3&   \phn34.2&   \phn36.1&\phn\phn1.55$\pm$0.65&\phn2.37&   \phn\phn6.4&$+0.62^{+0.36}_{-1.41}$&$-0.22^{+1.18}_{-0.76}$&$-0.74^{+0.98}_{-0.23}$&e\\
 7&CXOU J043138.3$-$050536&04 31 38.30&$-$05 05 36.7&   \phn35.2&   \phn39.2&\phn\phn7.48$\pm$1.35&\phn5.52&      \phn30.8&$+0.30^{+0.33}_{-0.42}$&$+0.12^{+0.38}_{-0.42}$&$-0.18^{+0.24}_{-0.22}$&\nodata\\
 8&CXOU J043141.2$-$050543&04 31 41.23&$-$05 05 43.0&   \phn38.4&   \phn49.1&\phn\phn4.77$\pm$1.10&\phn4.33&      \phn19.7&$-0.42^{+0.34}_{-0.25}$&$-0.67^{+0.34}_{-0.19}$&$-0.35^{+0.47}_{-0.34}$&e\\
 9&CXOU J043137.3$-$050457&04 31 37.33&$-$05 04 57.2&   \phn40.2&   \phn59.4&\phn\phn3.48$\pm$0.93&\phn3.73&      \phn14.3&$-0.26^{+0.41}_{-0.34}$&$-0.67^{+0.50}_{-0.23}$&$-0.50^{+0.57}_{-0.33}$&\nodata\\
10&CXOU J043139.8$-$050552&04 31 39.84&$-$05 05 52.3&   \phn41.9&   \phn43.4&\phn\phn8.32$\pm$1.42&\phn5.87&      \phn34.3&$-0.72^{+0.17}_{-0.11}$&$-0.86^{+0.17}_{-0.08}$&$-0.39^{+0.48}_{-0.33}$&f\\
11&CXOU J043137.3$-$050536&04 31 37.31&$-$05 05 36.0&   \phn46.0&   \phn56.6&\phn\phn1.43$\pm$0.61&\phn2.36&   \phn\phn5.9&$-0.72^{+1.05}_{-0.25}$&$-0.50^{+0.75}_{-0.37}$&$+0.35^{+0.59}_{-1.09}$&\nodata\\
12&CXOU J043141.9$-$050550&04 31 41.99&$-$05 05 50.8&   \phn51.3&   \phn68.0&\phn\phn2.03$\pm$0.73&\phn2.78&   \phn\phn8.4&$+0.25^{+0.50}_{-0.68}$&$-0.65^{+1.16}_{-0.32}$&$-0.78^{+0.93}_{-0.20}$&\nodata\\
13&CXOU J043141.5$-$050421&04 31 41.50&$-$05 04 21.0&   \phn55.1&   \phn56.3&\phn\phn1.52$\pm$0.63&\phn2.40&   \phn\phn6.3&$+0.06^{+0.62}_{-0.66}$&$-1.00^{+0.66}_{-0.00}$&$-1.00^{+0.59}_{-0.00}$&e\\
14&CXOU J043139.8$-$050407&04 31 39.81&$-$05 04 07.1&   \phn63.4&   \phn66.1&\phn\phn3.65$\pm$0.94&\phn3.90&      \phn15.0&$-0.15^{+0.42}_{-0.37}$&$-0.53^{+0.49}_{-0.29}$&$-0.40^{+0.48}_{-0.33}$&\nodata\\
15&CXOU J043136.0$-$050542&04 31 36.04&$-$05 05 42.2&   \phn65.5&   \phn83.2&\phn\phn1.66$\pm$0.65&\phn2.56&   \phn\phn6.9&$-0.27^{+0.91}_{-0.60}$&$+0.01^{+0.71}_{-0.72}$&$+0.28^{+0.51}_{-0.73}$&\nodata\\
16&CXOU J043135.1$-$050502&04 31 35.17&$-$05 05 02.6&   \phn70.7&      103.9&\phn\phn9.47$\pm$1.51&\phn6.26&      \phn39.0&$-0.25^{+0.22}_{-0.20}$&$-0.61^{+0.23}_{-0.16}$&$-0.43^{+0.27}_{-0.21}$&f\\
17&CXOU J043143.2$-$050411&04 31 43.27&$-$05 04 11.3&   \phn78.0&   \phn86.3&\phn\phn5.49$\pm$1.18&\phn4.63&      \phn22.6&$-0.48^{+0.24}_{-0.18}$&$-1.00^{+0.12}_{-0.00}$&$-1.00^{+0.33}_{-0.00}$&f\\
18&CXOU J043145.0$-$050511&04 31 45.09&$-$05 05 11.1&   \phn78.0&      113.5&\phn\phn1.66$\pm$0.65&\phn2.54&   \phn\phn6.8&$-0.47^{+0.59}_{-0.34}$&$-0.81^{+0.91}_{-0.17}$&$-0.54^{+1.12}_{-0.41}$&d\\
19&CXOU J043145.0$-$050438&04 31 45.06&$-$05 04 38.1&   \phn84.0&      111.0&\phn\phn1.24$\pm$0.56&\phn2.20&   \phn\phn5.1&$-0.35^{+0.87}_{-0.51}$&$-0.40^{+0.88}_{-0.47}$&$-0.06^{+0.77}_{-0.70}$&\nodata\\
20&CXOU J043139.9$-$050339&04 31 39.93&$-$05 03 39.2&   \phn91.3&   \phn94.6&\phn\phn1.48$\pm$0.60&\phn2.45&   \phn\phn6.1&$-0.02^{+0.89}_{-0.86}$&$+0.26^{+0.65}_{-1.00}$&$+0.28^{+0.51}_{-0.73}$&\nodata\\
21&CXOU J043134.2$-$050431&04 31 34.25&$-$05 04 31.1&   \phn92.7&      136.0&\phn\phn5.54$\pm$1.15&\phn4.82&      \phn22.8&$-0.72^{+0.22}_{-0.13}$&$-0.95^{+0.36}_{-0.04}$&$-0.73^{+1.00}_{-0.24}$&f\\
22&CXOU J043143.6$-$050627&04 31 43.65&$-$05 06 27.6&   \phn95.6&      125.5&\phn\phn3.75$\pm$0.97&\phn3.85&      \phn15.5&$+0.16^{+0.54}_{-0.65}$&$+0.04^{+0.59}_{-0.62}$&$-0.11^{+0.40}_{-0.37}$&\nodata\\
23&CXOU J043141.7$-$050336&04 31 41.78&$-$05 03 36.9&   \phn97.9&   \phn97.9&\phn\phn2.73$\pm$0.83&\phn3.30&      \phn11.2&$-0.47^{+0.67}_{-0.37}$&$-0.60^{+0.69}_{-0.30}$&$-0.18^{+0.75}_{-0.59}$&d\\
24&CXOU J043141.2$-$050654&04 31 41.25&$-$05 06 54.3&      105.8&      117.5&\phn\phn2.80$\pm$0.82&\phn3.43&      \phn11.5&$+0.01^{+0.61}_{-0.62}$&$-0.16^{+0.69}_{-0.56}$&$-0.17^{+0.51}_{-0.43}$&e\\
25&CXOU J043141.6$-$050653&04 31 41.61&$-$05 06 53.5&      106.3&      120.4&   \phn12.63$\pm$1.71&\phn7.38&      \phn52.0&$+0.41^{+0.25}_{-0.34}$&$+0.36^{+0.27}_{-0.34}$&$-0.05^{+0.17}_{-0.17}$&f\\
26&CXOU J043132.6$-$050513&04 31 32.61&$-$05 05 13.1&      108.5&      156.9&\phn\phn1.52$\pm$0.62&\phn2.46&   \phn\phn6.3&$-0.61^{+0.83}_{-0.32}$&$-0.59^{+0.73}_{-0.31}$&$+0.03^{+0.76}_{-0.80}$&\nodata\\
27&CXOU J043139.5$-$050711&04 31 39.58&$-$05 07 11.5&      121.1&      124.6&   \phn16.02$\pm$1.92&\phn8.32&      \phn66.0&$-0.29^{+0.16}_{-0.15}$&$-0.52^{+0.16}_{-0.13}$&$-0.27^{+0.18}_{-0.17}$&f\\
28&CXOU J043132.1$-$050428&04 31 32.13&$-$05 04 28.6&      123.0&      181.5&\phn\phn3.31$\pm$0.89&\phn3.71&      \phn13.6&$-0.46^{+0.33}_{-0.24}$&$-0.83^{+0.38}_{-0.13}$&$-0.60^{+0.63}_{-0.29}$&\nodata\\
29&CXOU J043144.8$-$050331&04 31 44.80&$-$05 03 31.9&      123.1&      132.8&   \phn11.38$\pm$1.64&\phn6.92&      \phn46.9&$-0.22^{+0.19}_{-0.17}$&$-0.68^{+0.19}_{-0.13}$&$-0.54^{+0.22}_{-0.17}$&f\\
30&CXOU J043132.2$-$050422&04 31 32.25&$-$05 04 22.6&      123.5&      181.7&   \phn13.46$\pm$1.82&\phn7.38&      \phn55.4&$+0.30^{+0.24}_{-0.28}$&$+0.05^{+0.29}_{-0.30}$&$-0.26^{+0.18}_{-0.16}$&e,f\\
31&CXOU J043147.5$-$050612&04 31 47.55&$-$05 06 12.6&      130.5&      190.2&\phn\phn0.68$\pm$0.43&\phn1.56&   \phn\phn2.8&$-1.00^{+0.60}_{-0.00}$&$-1.00^{+0.44}_{-0.00}$&$+0.00^{+1.00}_{-1.00}$&\nodata\\
32&CXOU J043147.1$-$050353&04 31 47.18&$-$05 03 53.5&      133.6&      163.0&\phn\phn3.34$\pm$0.91&\phn3.69&      \phn13.8&$-0.16^{+0.48}_{-0.42}$&$-0.31^{+0.51}_{-0.38}$&$-0.15^{+0.44}_{-0.39}$&\nodata\\
33&CXOU J043136.1$-$050308&04 31 36.19&$-$05 03 08.3&      133.9&      162.7&\phn\phn1.71$\pm$0.65&\phn2.65&   \phn\phn7.1&$-0.52^{+0.63}_{-0.33}$&$-0.81^{+0.91}_{-0.17}$&$-0.50^{+1.13}_{-0.45}$&d\\
34&CXOU J043146.4$-$050338&04 31 46.48&$-$05 03 38.4&      135.1&      156.1&\phn\phn2.00$\pm$0.71&\phn2.83&   \phn\phn8.2&$-0.43^{+0.59}_{-0.37}$&$-0.81^{+0.91}_{-0.17}$&$-0.58^{+1.11}_{-0.38}$&\nodata\\
35&CXOU J043142.2$-$050257&04 31 42.24&$-$05 02 57.2&      137.8&      137.8&\phn\phn1.26$\pm$0.56&\phn2.25&   \phn\phn5.2&$+0.31^{+0.60}_{-1.03}$&$-0.40^{+1.24}_{-0.56}$&$-0.64^{+1.08}_{-0.33}$&\nodata\\
36&CXOU J043131.3$-$050411&04 31 31.34&$-$05 04 11.8&      140.2&      205.8&\phn\phn2.97$\pm$0.87&\phn3.40&      \phn12.2&$-0.59^{+0.40}_{-0.23}$&$-0.96^{+1.88}_{-0.04}$&$-0.86^{+1.84}_{-0.14}$&\nodata\\
37&CXOU J043141.8$-$050252&04 31 41.84&$-$05 02 52.5&      141.1&      141.3&\phn\phn1.93$\pm$0.68&\phn2.82&   \phn\phn8.0&$+0.41^{+0.57}_{-1.29}$&$+0.40^{+0.58}_{-1.28}$&$-0.01^{+0.58}_{-0.57}$&\nodata\\
38&CXOU J043148.6$-$050415&04 31 48.60&$-$05 04 15.9&      141.5&      186.8&\phn\phn1.33$\pm$0.59&\phn2.27&   \phn\phn5.5&$-0.35^{+0.87}_{-0.51}$&$-0.40^{+0.88}_{-0.47}$&$-0.06^{+0.77}_{-0.70}$&\nodata\\
39&CXOU J043138.2$-$050245&04 31 38.21&$-$05 02 45.4&      147.2&      161.8&\phn\phn1.25$\pm$0.56&\phn2.23&   \phn\phn5.2&$+0.57^{+0.42}_{-1.39}$&$-1.00^{+2.00}_{-0.00}$&$-1.00^{+0.55}_{-0.00}$&\nodata\\
40&CXOU J043149.8$-$050539&04 31 49.86&$-$05 05 39.7&      152.2&      224.8&\phn\phn3.16$\pm$0.87&\phn3.62&      \phn13.0&$-1.00^{+0.51}_{-0.00}$&$-0.67^{+0.57}_{-0.24}$&$+1.00^{+0.00}_{-2.00}$&d\\
41&CXOU J043147.0$-$050308&04 31 47.07&$-$05 03 08.7&      162.5&      180.7&\phn\phn2.88$\pm$0.84&\phn3.45&      \phn11.9&$-0.16^{+0.48}_{-0.42}$&$-0.52^{+0.58}_{-0.32}$&$-0.39^{+0.55}_{-0.37}$&\nodata\\
42&CXOU J043132.2$-$050712&04 31 32.25&$-$05 07 12.0&      166.6&      187.2&\phn\phn4.88$\pm$1.13&\phn4.31&      \phn20.1&$+1.00^{+0.00}_{-1.47}$&$+1.00^{+0.00}_{-0.94}$&$+0.22^{+0.30}_{-0.35}$&f\\
43&CXOU J043148.8$-$050328&04 31 48.83&$-$05 03 28.1&      168.5&      202.3&\phn\phn1.10$\pm$0.52&\phn2.09&   \phn\phn4.5&$-0.35^{+1.21}_{-0.61}$&$+0.12^{+0.76}_{-0.93}$&$+0.45^{+0.49}_{-1.10}$&\nodata\\
44&CXOU J043146.0$-$050735&04 31 46.08&$-$05 07 35.7&      172.4&      221.3&\phn\phn1.30$\pm$0.60&\phn2.16&   \phn\phn5.3&$-0.21^{+0.67}_{-0.52}$&$-1.00^{+0.21}_{-0.00}$&$-1.00^{+0.31}_{-0.00}$&\nodata\\
45&CXOU J043135.5$-$050222&04 31 35.50&$-$05 02 22.7&      180.0&      214.2&\phn\phn1.76$\pm$0.67&\phn2.65&   \phn\phn7.30&$+0.46^{+0.47}_{-1.07}$&$-0.40^{+1.24}_{-0.56}$&$-0.73^{+1.00}_{-0.24}$&\nodata\\
46&CXOU J043127.8$-$050537&04 31 27.82&$-$05 05 37.5&      182.0&      257.3&\phn\phn6.80$\pm$1.29&\phn5.28&      \phn28.0&$-0.04^{+0.37}_{-0.36}$&$-0.04^{+0.36}_{-0.35}$&$-0.00^{+0.28}_{-0.28}$&f\\
47&CXOU J043142.9$-$050213&04 31 42.98&$-$05 02 13.3&      183.2&      183.2&\phn\phn1.90$\pm$0.89&\phn2.14&   \phn\phn7.8&$+1.00^{+0.00}_{-2.00}$&$+1.00^{+0.00}_{-2.00}$&$-0.26^{+0.77}_{-0.54}$&c\\
48&CXOU J043137.8$-$050818&04 31 37.80&$-$05 08 18.3&      190.4&      191.5&\phn\phn3.12$\pm$0.89&\phn3.53&      \phn12.9&$-0.65^{+0.50}_{-0.24}$&$-0.74^{+0.50}_{-0.19}$&$-0.18^{+0.75}_{-0.59}$&\nodata\\
49&CXOU J043127.7$-$050405&04 31 27.75&$-$05 04 05.9&      192.2&      283.9&\phn\phn1.91$\pm$0.69&\phn2.77&   \phn\phn7.93&$+0.21^{+0.53}_{-0.68}$&$-0.28^{+0.80}_{-0.54}$&$-0.46^{+0.59}_{-0.35}$&e\\
50&CXOU J043130.4$-$050256&04 31 30.44&$-$05 02 56.1&      194.6&      269.3&   \phn75.83$\pm$4.18&   18.14&         312.4&$-0.78^{+0.03}_{-0.03}$&$-0.99^{+0.02}_{-0.01}$&$-0.92^{+0.12}_{-0.05}$&d\\
51&CXOU J043150.1$-$050304&04 31 50.16&$-$05 03 04.7&      198.7&      235.5&   \phn23.13$\pm$2.34&\phn9.90&      \phn95.3&$-0.54^{+0.10}_{-0.09}$&$-0.77^{+0.09}_{-0.07}$&$-0.39^{+0.18}_{-0.16}$&f\\
52&CXOU J043134.2$-$050205&04 31 34.21&$-$05 02 05.2&      203.7&      247.9&\phn\phn5.99$\pm$2.74&\phn2.21&      \phn24.7&$+1.00^{+0.00}_{-2.00}$&$+1.00^{+0.00}_{-2.00}$&$+0.56^{+0.39}_{-1.09}$&c\\
53&CXOU J043147.4$-$050212&04 31 47.45&$-$05 02 12.5&      210.9&      221.9&      116.05$\pm$7.08&   16.41&         478.1&$-0.41^{+0.06}_{-0.06}$&$-0.76^{+0.05}_{-0.04}$&$-0.50^{+0.08}_{-0.08}$&c,f\\
54&CXOU J043134.6$-$050826&04 31 34.68&$-$05 08 26.8&      211.0&      212.6&\phn\phn4.45$\pm$1.10&\phn4.05&      \phn18.3&$+0.04^{+0.38}_{-0.40}$&$-0.65^{+0.58}_{-0.25}$&$-0.67^{+0.50}_{-0.23}$&\nodata\\
55&CXOU J043153.6$-$050408&04 31 53.68&$-$05 04 08.6&      215.4&      293.8&   \phn26.29$\pm$2.64&\phn9.94&         108.3&$-0.26^{+0.13}_{-0.12}$&$-0.61^{+0.12}_{-0.10}$&$-0.42^{+0.15}_{-0.13}$&f\\
56&CXOU J043150.9$-$050730&04 31 50.96&$-$05 07 30.7&      217.1&      304.9&\phn\phn2.77$\pm$0.86&\phn3.24&      \phn11.4&$-0.13^{+0.47}_{-0.42}$&$-0.72^{+0.71}_{-0.23}$&$-0.65^{+0.73}_{-0.27}$&\nodata\\
57&CXOU J043125.4$-$050540&04 31 25.43&$-$05 05 40.6&      217.8&      308.5&\phn\phn4.11$\pm$1.04&\phn3.96&      \phn17.0&$-0.04^{+0.35}_{-0.34}$&$-0.79^{+0.58}_{-0.17}$&$-0.78^{+0.57}_{-0.18}$&\nodata\\
58&CXOU J043126.0$-$050639&04 31 26.07&$-$05 06 39.6&      224.6&      295.4&\phn\phn5.96$\pm$1.24&\phn4.79&      \phn24.6&$+0.61^{+0.36}_{-1.20}$&$+0.74^{+0.24}_{-1.13}$&$+0.24^{+0.26}_{-0.30}$&f\\
59&CXOU J043153.0$-$050659&04 31 53.06&$-$05 06 59.0&      225.0&      327.7&\phn\phn4.65$\pm$1.07&\phn4.35&      \phn19.2&$-0.51^{+0.27}_{-0.20}$&$-0.78^{+0.30}_{-0.13}$&$-0.46^{+0.50}_{-0.32}$&e,f\\
60&CXOU J043153.9$-$050656&04 31 53.90&$-$05 06 56.8&      235.0&      343.6&\phn\phn4.29$\pm$1.05&\phn4.09&      \phn17.7&$+0.69^{+0.30}_{-1.42}$&$+0.76^{+0.24}_{-1.41}$&$+0.13^{+0.30}_{-0.33}$&\nodata\\
61&CXOU J043131.2$-$050832&04 31 31.21&$-$05 08 32.6&      240.0&      252.8&\phn\phn2.40$\pm$0.84&\phn2.86&   \phn\phn9.9&$+0.56^{+0.39}_{-1.07}$&$-0.17^{+1.03}_{-0.76}$&$-0.66^{+0.81}_{-0.28}$&\nodata\\
62&CXOU J043123.9$-$050546&04 31 23.95&$-$05 05 46.2&      240.6&      340.1&\phn\phn2.22$\pm$0.78&\phn2.84&   \phn\phn9.1&$+1.00^{+0.00}_{-1.95}$&$+1.00^{+0.00}_{-1.55}$&$+0.11^{+0.44}_{-0.49}$&\nodata\\
63&CXOU J043126.5$-$050729&04 31 26.51&$-$05 07 29.7&      243.4&      297.4&\phn\phn3.62$\pm$1.00&\phn3.62&      \phn14.9&$-0.46^{+0.71}_{-0.39}$&$-0.67^{+0.96}_{-0.29}$&$-0.31^{+0.99}_{-0.59}$&d\\
64&CXOU J043155.4$-$050311&04 31 55.43&$-$05 03 11.5&      261.2&      335.9&\phn\phn6.25$\pm$1.27&\phn4.92&      \phn25.7&$-0.08^{+0.30}_{-0.29}$&$-0.60^{+0.35}_{-0.21}$&$-0.55^{+0.34}_{-0.22}$&f\\
65&CXOU J043155.9$-$050255&04 31 55.98&$-$05 02 55.1&      276.2&      350.1&\phn\phn6.53$\pm$1.31&\phn5.00&      \phn26.9&$-0.03^{+0.31}_{-0.31}$&$-0.51^{+0.37}_{-0.25}$&$-0.48^{+0.32}_{-0.23}$&f\\
66&CXOU J043124.2$-$050232&04 31 24.22&$-$05 02 32.3&      282.3&      405.2&\phn\phn0.82$\pm$0.47&\phn1.76&   \phn\phn3.4&$+1.00^{+0.00}_{-2.00}$&$+1.00^{+0.00}_{-2.00}$&$-0.80^{+1.79}_{-0.20}$&\nodata\\
67&CXOU J043130.2$-$050918&04 31 30.29&$-$05 09 18.3&      286.1&      297.3&   \phn14.34$\pm$1.99&\phn7.21&      \phn59.1&$-0.12^{+0.19}_{-0.18}$&$-0.67^{+0.23}_{-0.15}$&$-0.60^{+0.24}_{-0.17}$&f\\
68&CXOU J043138.6$-$051011&04 31 38.65&$-$05 10 11.1&      301.2&      308.2&\phn\phn6.14$\pm$1.35&\phn4.54&      \phn25.3&$-0.45^{+0.27}_{-0.21}$&$-0.64^{+0.29}_{-0.18}$&$-0.27^{+0.41}_{-0.33}$&f\\
69&CXOU J043150.8$-$050953&04 31 50.83&$-$05 09 53.3&      326.8&      412.8&\phn\phn4.10$\pm$1.17&\phn3.50&      \phn16.9&$+0.51^{+0.45}_{-1.22}$&$-0.65^{+1.63}_{-0.35}$&$-0.87^{+1.75}_{-0.13}$&\nodata\\
70&CXOU J043127.7$-$051021&04 31 27.75&$-$05 10 21.0&      359.5&      373.9&   \phn43.34$\pm$3.52&   12.32&         178.6&$+0.14^{+0.12}_{-0.12}$&$-0.39^{+0.13}_{-0.12}$&$-0.51^{+0.09}_{-0.08}$&f\\
71&CXOU J043157.3$-$051006&04 31 57.39&$-$05 10 06.5&      395.1&      533.9&   \phn13.61$\pm$3.63&\phn3.91&      \phn56.1&$+0.04^{+0.33}_{-0.34}$&$-0.67^{+0.44}_{-0.22}$&$-0.69^{+0.37}_{-0.19}$&c,e\\
\enddata
\tablecomments{The units for $L_X$ are $10^{38}$ erg s$^{-1}$ in the
0.3--10~keV band.}
\tablenotetext{a}{Sources near the center may be confused with
nearby sources, making their positions, fluxes, and extents uncertain.}
\tablenotetext{b}{Source is noticeably more extended than PSF.}
\tablenotetext{c}{Source is at the edge of the S3 detector, and
flux is uncertain due to large exposure correction.}
\tablenotetext{d}{Possible optical counterpart.}
\tablenotetext{e}{Source may be variable.}
\tablenotetext{f}{Source is part of analysis sample.}
\end{deluxetable}
\clearpage
\end{landscape}

\clearpage

\pagestyle{empty}

\begin{landscape}
\tabletypesize{\scriptsize}
\begin{deluxetable}{llcclccccccrc}
\tabletypesize{\scriptsize}
\tablecaption{X-ray Spectral Fits of NGC~1600 \label{tab:spectra_n1600}}
\tablehead{
&&&&
\multicolumn{3}{c}{Hard Component}&&
\multicolumn{3}{c}{Soft Component (mekal)}&&\\
\cline{5-7} \cline{9-11}
&&&
\colhead{$N_H$}&
\colhead{Model}&
\colhead{$kT_h$ or $\Gamma$}&
\colhead{$F^h_X$}&&
\colhead{$kT_s$}&
\colhead{Abund.}&
\colhead{$F^s_X$)}&&\\
\colhead{Row}&
\colhead{Origin}&
\colhead{Region}&
\colhead{($10^{20}$ cm$^{-2}$)}&&
\colhead{(keV)}&
\colhead{(\tablenotemark{a} \,)}&&
\colhead{(keV)}&
\colhead{(solar)}&
\colhead{(\tablenotemark{a} \,)}&
\colhead{Counts}&
\colhead{$\chi^2$/dof}}
\startdata
1                 &Sources   &Field              &(4.86)                    &Bremss&$4.73^{+1.24}_{-0.89}$&2.78&&\nodata&\nodata&\nodata&   1318&\phn\phn   52.2/48=1.09\\
2                 &Sources   &Field              &0.48 [$<$6.32]            &Bremss&$5.69^{+1.88}_{-1.57}$&2.75&&\nodata&\nodata&\nodata&   1318&\phn\phn   50.7/47=1.08\\
3\tablenotemark{b}&Sources   &Field              &(4.86)                    &Power &$1.76^{+0.10}_{-0.09}$&3.26&&\nodata&\nodata&\nodata&   1318&\phn\phn   52.4/48=1.09\\
4                 &Sources   &Field              &$8.62^{+7.60}_{-7.74}$    &Power &$1.84^{+0.20}_{-0.18}$&3.35&&\nodata&\nodata&\nodata&   1318&\phn\phn   51.8/47=1.10\\
5                 &Sources   &Field              &(4.86)                    &Diskbb&1.44 [$\ge$0.00]          &0.88&&\nodata&\nodata&\nodata&   1318&\phn\phn   50.5/46=1.10\\*
                  &          &                   &(4.86)                    &Power &$1.95^{+2.23}_{-0.46}$&2.10&&\nodata&\nodata&\nodata&       &\\
6\tablenotemark{c}   &Unresolved&1 $a_{\rm eff}$           &(4.86)                    &\nodata&\nodata                   &\nodata               &&$1.02^{+0.03}_{-0.03}$&$0.15^{+0.03}_{-0.02}$&4.95&3662   &\phn   168.6/87 =1.94\\
7\tablenotemark{c}   &Unresolved&1 $a_{\rm eff}$           &(4.86)                    &Power  &(1.76)                    &$2.29^{+0.47}_{-0.48}$&&$0.95^{+0.03}_{-0.03}$&$0.27^{+0.11}_{-0.07}$&3.46&3662   &\phn   110.0/86 =1.27\\
8\tablenotemark{c}   &Unresolved&1 $a_{\rm eff}$           &$7.57^{+7.12}_{-4.61}$    &Power  &(1.76)                    &$3.42^{+0.72}_{-0.73}$&&$0.93^{+0.06}_{-0.06}$&$0.26^{+0.10}_{-0.06}$&3.74&3662   &\phn   108.8/85 =1.28\\
9\tablenotemark{c}   &Unresolved&1 $a_{\rm eff}$           &(4.86)                    &Power  &(1.76)                    &0.07 [$<$1.76]        &&$0.84^{+0.03}_{-0.04}$&1000 [$>$0.59]        &1.70&3662   &\phn\phn78.7/83 =0.95\\*
                     &          &                          &                          &       &                          &                      &&$2.38^{+0.44}_{-0.74}$&$0.79^{+0.72}_{-0.37}$&3.25&&\\
10\tablenotemark{c,d}&Unresolved&1 $a_{\rm eff}$           &(4.86)                    &Power  &(1.76)                    &0.19 [$<$1.93]        &&$0.85^{+0.04}_{-0.04}$&$1.07^{+1.00}_{-0.40}$&2.04&3662   &\phn\phn79.9/84 =0.95\\*
                     &          &                          &                          &       &                          &                      &&$2.55^{+0.52}_{-0.86}$&$1.07^{+1.00}_{-0.40}$&3.25&&\\
11\tablenotemark{c}  &Unresolved& $a<9\arcsec$             &(4.86)                    &Power  &(1.76)                    &$0.59^{+0.23}_{-0.23}$&&$0.84^{+0.04}_{-0.04}$&$0.46^{+1.29}_{-0.22}$&1.04&1078   &\phn\phn24.4/28=0.87\\
12\tablenotemark{c}  &Unresolved& $9\arcsec<a<18\arcsec$   &(4.86)                    &Power  &(1.76)                    &$0.57^{+0.19}_{-0.20}$&&$0.88^{+0.05}_{-0.04}$&$0.64^{+5.46}_{-0.33}$&0.89&1010   &\phn\phn27.8/28=0.99\\
13\tablenotemark{c}  &Unresolved& $18\arcsec<a<41\arcsec$  &(4.86)                    &Power  &(1.76)                    &$0.83^{+0.37}_{-0.41}$&&$1.29^{+0.18}_{-0.13}$&$0.35^{+0.41}_{-0.16}$&0.90&\phn981&\phn\phn45.4/36=1.26\\
14\tablenotemark{c}  &Unresolved& $41\arcsec<a<69\arcsec$  &(4.86)                    &Power  &(1.76)                    &0.30 [$<$0.77]        &&$1.43^{+0.21}_{-0.16}$&$0.31^{+0.27}_{-0.14}$&1.20&\phn939&\phn\phn51.2/42=1.22\\
15\tablenotemark{c}  &Unresolved& $69\arcsec<a<98\arcsec$  &(4.86)                    &Power  &(1.76)                    &$1.52^{+0.24}_{-0.24}$&&$1.48^{+0.24}_{-0.17}$&995.21 [$>$0.788]     &0.41&\phn976&\phn\phn48.7/54=0.90\\
16\tablenotemark{c}  &Unresolved& $98\arcsec<a<126\arcsec$ &(4.86)                    &Power  &(1.76)                    &$0.74^{+0.60}_{-0.72}$&&$1.62^{+0.27}_{-0.19}$&$0.86^{+8.62}_{-0.42}$&0.44&\phn961&\phn\phn57.4/60=0.96\\
17\tablenotemark{c}  &Unresolved& $126\arcsec<a<155\arcsec$&(4.86)                    &Power  &(1.76)                    &$1.09^{+0.61}_{-0.70}$&&$1.38^{+0.35}_{-0.22}$&$0.31^{+1.67}_{-0.19}$&0.71&\phn978&\phn\phn78.6/71=1.11\\
18\tablenotemark{c}  &Unresolved& $155\arcsec<a<180\arcsec$&(4.86)                    &Power  &(1.76)                    &0.05 [$<$0.75]        &&$1.63^{+0.36}_{-0.28}$&$0.31^{+0.46}_{-0.17}$&1.11&\phn778&\phn\phn76.3/65=1.17\\
\enddata
\tablenotetext{a}{Units are $10^{-13}$ erg cm$^{-2}$ s$^{-1}$ in 0.3--10
keV band.}
\tablenotetext{b}{The adopted best-fit model for this emission.}
\tablenotetext{c}{The energy range for this spectrum excludes 1.9 - 2.1~keV.}
\tablenotetext{d}{The abundances for this spectrum were tied together.}
\end{deluxetable}
\clearpage
\end{landscape}

\end{document}